\documentclass[letterpaper, 11pt]{article} 

\pdfoutput=1
\usepackage{times,abstract}
\usepackage[authoryear]{natbib}
\usepackage[pdftex]{color,graphicx} 
\usepackage{amsxtra}     
\usepackage{amsthm,gensymb}
\usepackage{amssymb}
\usepackage{amsfonts}
\usepackage{upgreek}
\usepackage{subcaption}
\usepackage{array} 
\usepackage{lscape}
\usepackage{rotating} 
\usepackage[table]{xcolor}
\usepackage{cellspace}
\usepackage{amsthm} 
\usepackage{verbatim}  
\usepackage{graphicx} 
\usepackage{amsmath} 
\usepackage{amssymb} 
\usepackage{float}
\usepackage{multirow}
\usepackage{array} 
\usepackage{ragged2e}
\usepackage{picinpar} 
\usepackage{rotating} 
\usepackage{fancybox} 
\usepackage{wrapfig} 
\usepackage[all,2cell]{xy} 
\usepackage{epigraph}
\usepackage{grffile}
\usepackage{soul}
\usepackage[bf]{titlesec}
\usepackage{setspace}    
\UseAllTwocells
\theoremstyle{plain}

\theoremstyle{remark}

\theoremstyle{definition}

\renewcommand{\i}[1]{\textit{#1}}

\definecolor{tableShade}{HTML}{F1F1F1}

\titleformat{\section}{\large\bfseries}{\thesection.}{1mm}{}
\titleformat{\subsection}{\normalsize\bfseries}{\thesubsection}{1mm}{}
\newcolumntype{V}{>{\centering\arraybackslash} m{.4\linewidth} }

\numberwithin{equation}{section}

\makeatletter
\let\@fnsymbol\@arabic
\makeatother

\newcommand{\affiliation}[2]{%
   \footnote{#2}
    \newcounter{#1}
    \setcounter{#1}{\value{footnote}}%
}
\newcommand{\sameaffiliation}[1]{%
    \footnotemark[\value{#1}]%
} 

\newlength{\imgwidth}

\begin{document}

\bibliographystyle{plainnat}    
\bibpunct{(}{)}{,}{a}{}{;}

\title{Dynamical Properties of Interaction Data}
\author{Aaron Bramson\affiliation{Riken}{Laboratory for Symbolic Cognitive Development, Riken Brain Science Institute} \affiliation{Gent-Econ}{Department of General Economics, Ghent University} \affiliation{UNCC}{Department of Software and Information Systems, University of North Carolina at Charlotte} 
\and Benjamin Vandermarliere \sameaffiliation{Gent-Econ} \hspace{0.2mm} \affiliation{Gent-Physics}{Department of Physics and Astronomy, Ghent University}%
}

\singlespace

  \maketitle             
  \begin{onecolabstract} 
  Network dynamics are typically presented as a time series of network properties captured at each period. The current approach examines the dynamical properties of transmission via novel measures on an integrated, temporally extended network representation of interaction data across time. Because it encodes time and interactions as network connections, static network measures can be applied to this ``temporal web'' to reveal features of the dynamics themselves. Here we provide the technical details and apply it to agent-based implementations of the well-known SEIR and SEIS epidemiological models.  
\vspace{10mm}
  \end{onecolabstract}

\section{Introduction}\label{Introduction}

Network measures provide useful insight into the structure of relationships and interactions, and the breadth of systems that can be represented as a network has fostered an explosive growth in network analyses across all disciplines (for an introduction see \citep{newman2010networks}).  Separately, analyzing system dynamics is common to all sciences; sometimes as differential equations, other times as regressions on time series data, and yet other times in animations or sequences of large patterns in data (e.g., changes in spatial maps or structural diagrams).  But these standard approaches capture only the dynamics of a measure rather than a true measure of the system's dynamics.  In order to measure a system's processes directly, the methodology presented here captures the interaction and/or structural dynamics in a temporally extended network representation. Network measures applied to this ``temporal web'' then reveal features of the processes itself. 

A number of recent papers have outlined similar techniques for capturing dynamic networks and/or dynamics on networks using a layered graph structure \citep{Holme2012,Nicosia2013}.  Although these other temporal graphs have layers through time, those layers are not connected and the analysis focuses on time slices of the graph.  Using the standard or adapted network measures for structural properties, values for time slices are calculated and compared to other slices.  Our approach utilizes connections across time, i.e., spanning multiple slices.  While standard ``temporal networks are not graphs'' \citep{Lentz2013}, but rather series of graphs in temporal layers, our technique creates a single graph through transtemporal edges.  We further differ in that our analytical technique is applied to this temporally extruded graph (i.e., the graph generated by connecting the layers) and its cross-temporal subgraphs rather than to time-slice layers.  This makes a very different class of measures useful and provides distinct insights into the system's dynamics. 

To demonstrate the technique we apply it to agent-based models (ABM) of both SEIR and SEIS epidemiological systems (i.e., a disease for which people go through the stages susceptible, exposed, infectious, and removed/recovered and alternatively in which people become susceptible again after the infectious state).  This class of models offers a good test case because it has been well explored and is analogous to processes in many fields outside epidemiology (spread of ideas, technology, ...) \citep{Newman2002}.  Using an ABM instead of a categorical or stochastic differential equation model allows us to record the actual transmission events as well as perform agent- and time-specific contingency analyses.  The purpose of this paper is not to contribute substantively to the modeling of SEIR or SEIS dynamics, but rather to demonstrate a new technique for measuring those dynamics. For this reason we have calibrated our SEIR ABM models to the ones presented by Rahmandad and Sterman \citep{Rahmandad2008}.  By doing so we intend to inherit their description and minimize our exposition of the ABM model, its comparison to differential equations models, and the implications of heterogeneity in the ABM for health policy and transmission effects.  Our SEIS model uses the same parameters as the SEIR model, and furthermore is build using the same interaction ``skeleton'' and random seeds as the SEIR models to provide the closest comparisons (more details below).


The purpose here is not to endorse a particular form of modeling or to contribute substantively to epidemiology in a direct way.  The purpose is to present a \i{methodology for examining intertemporal interaction data}, and the agent-based modeling approach of this epidemiology problem is convenient for generating data with the appropriate structure.   It also provides a clear demonstration that the method produces additional insight into an already well understood phenomenon.  The key is the interaction data.  Upcoming research utilizes data collected from transactions among banks, neural activity (connectome) data across brain regions, inventory flows in logistic systems, and traffic patterns on road-rail-flight networks.


Techniques such as this one are important for delivering the potential of computer simulations of complex adaptive systems.  For example, the cornerstone concept of \i{emergence} is purported to be a property of a system's interactions and dynamics, and therefore formal definitions of emergence may become available by analyzing the temporal webs of such systems.  We do not attempt that task here, but instead make progress on current methodologies' inadequate ability to capture dynamics and analyze processes.  For example, the projection or flattening of the interaction data into a single-layered graph makes the standard collection of network measures applicable, but it produces unreliable and/or incorrect propagation dynamics \citep{Pfitzner2013}.  Our evaluation of the usefulness of existing measures (such as diameter, clustering, betweenness, eigenvector centrality, ...) on the temporal web revealed that these intuitive and useful time-slice measures fail to provide useful information on the temporally extended network representation (for reasons explained below).  We therefore propose a new suite of measures grounded in network flow, in- and out-components, time-reversed in- and out-components, and transtemporal motifs.  We demonstrate the effectiveness of this class of measures for identifying the agents \i{and times} upon which disease spread is the most contingent using a comparison to an exhaustive temporal knockout (TKO) measure.


\section{The Agent-Based SEIR and SEIS Models}

Our agent-based model of SEIR disease progression and spread follows the specification in Rahmandad and Sterman \citep{Rahmandad2008}.\footnote{We provide the core details here for ease of reference, which will suffice for those already familiar with SEIR models.  For more details about their ABM model, and their comparison to differential equations models, please refer to their paper.}  For the SEIS we used an identical setup and parameters, only changing the I$\rightarrow$R transition to I$\rightarrow$S.

Our simulations contain 200 agents that are connected in an explicit undirected base network -- the collection of potential interaction conduits.  
The probability that agent $i$ contacts agent $j$, given they're connected in the base network, is: 
$$ c_{ij} = \frac{ \frac{1}{k_j}}{ \Sigma_{K_i} \frac{1}{ k_n } }$$
 with $k_j$ being the undirected degree of agent $j$, and the summation in the denominator is over each network neighbor ($n$) of node $i$ (written $K_i$).  For the results presented here we include only a fully connected \i{base} network which makes agent interaction uniformly random over the full set of other agents. In our follow-up research we also analyze random, small world, and scale-free base networks following Rahmandad and Sterman \citep{Rahmandad2008}, but we have omitted those scenarios in order to maintain focus on presenting the novel analysis technique being described in this paper.
 Updates are performed synchronously, so that that each agent's state at $t$ depends on the states of agents at $t-1$.  For each trial, these contact and state-changing dynamics are run for 400 time steps in order to ensure most SEIR infections are able to run their course and most SEIS infections reach full penetration. 
  
Each individual has four key parameters: (1) expected contact rate, (2) infectivity, (3) emergence time, and (4) disease duration.  The values for these characteristics (below) are calibrated to statistics for the common cold according to Rahmandad and Sterman \citep{Rahmandad2008}.  Agents are in one of four states: susceptible, exposed, infectious, or recovered. 
Disease spread dynamics follow these rules:

\begin{itemize}
\item Initially two agents (1\% of the population) are chosen uniformly at random and set to the infectious state.
\item If an exposed agent contacts a susceptible agent, then the latter has a probability $i_{SE} = 0.05$ to become exposed (i.e., enter the first stage of the disease).
\item If an infected agent contacts a susceptible agent, then the latter has a probability $i_{SI} = 0.06$ to become exposed.\footnote{In this model the exposed state can be seen as an infectious but non-symptomatic state, whereas during the infectious state the agent shows symptoms. In many SEIR models the exposed agents are considered to be in an incubation stage and therefore not able to infect other agents (which matches the labels of the states better).  In those models the time spent in the exposed state is referred to as a latency period between initial infection and overt illness.  Thus the Rahmandad and Sterman model can be seen as an S-I$_{1}$-I$_{2}$-R model if one prefers that nomenclature.  The distinction is not an important difference for our purposes here.}
\item At each time step, exposed agents have a probability of $1/15 = 0.066\bar{6}$ to enter the next stage of the disease, and become infectious.
\item Infectious agents likewise have a probability of $1/15$ to become recovered in the SEIR case, or to become susceptible in the SEIS case.
\item Recovered agents stay in that state indefinitely, and so they can also be thought of as removed.
\end{itemize}

\section{Building a Temporal Web}

The measures of dynamical properties utilized in this paper depend on a specific temporally extended network representation, so we now detail how to construct this representation.  The ``temporal web'' presented here is distinct from other ``temporal graphs'', ``temporal networks'', or ``layered networks'' in its connections across time; however, they are all variations on a theme.  As this subfield matures a better nomenclature will become necessary to disambiguate the techniques, although we will not propose one here.

The first step in building a temporal web is to decide upon a property to track: for us here it is the disease state.  In our example we generate a temporal web directly from the model dynamics described above, but the technique is not a way to build generative models. Typically one will feed the appropriate interaction and state-change data (either empirical data or from a model's output) into a post-processing temporal web generating and analysis routine (for which we plan to offer a python package in the future).  

In other applications, the original model or dataset may include a tremendous level of detail.  Our SEIR system only has one property of interest, but separate temporal webs can be constructed from the same system for each property that one wishes to analyze in this way.   These interactions may represent any sort of relationship among those elements: e.g., physical proximity, sharing of an idea, flow of resources, level of attraction, social obligations, financial debt, or any of the myriad relations that have been (or could be) encoded in a network. The nodes may likewise encode any properties that (potentially) change in response to what is chosen to be represented as the edges.  


The only requirements for building a temporal web from a dataset is that it includes both: 
\begin{itemize}
\item State changes in the elements across time and 
\item Interactions and/or relationships among the elements (e.g. agents) indexed by time. 
\end{itemize} 
Each period in the dataset becomes a layer in the temporal web; every element at each period is represented as a node in that layer.\footnote{An expansion of the technique under development allows for continuous time and heterogeneous time intervals (i.e., event-driven dynamics).}  The first period of state data is assigned to $t=0$ and captures the initial values for the nodes.  For each element, the node representing that element at $t$ is connected to the node for the same element at period $t+1$ (if it exists) with a directed edge following the flow of time.  These ``temporal edges'' connect the layers through the assumption that every element interacts with, or is related to, its former self.  Naturally this assumption can be relaxed when appropriate.

Network connections among the elements (such as social connections, transactions, spatial relations, etc.) exist at the layer(s) representing when they occur.  If the edges represent an existing \i{relationship} (directed or undirected) among the elements, then those relationships will usually be attached between pairs of nodes at the \i{same time step} (intra-temporal).  However, if the edges represent an \i{interaction, behavior, or process}, then that often happens across time and the nodes should be linked with a directed edge \i{across time} accordingly (cross-temporal).  It is possible to have both flavors of inter-personal edges (intra-temporal and cross-temporal) in the same temporal web (e.g., ideas can spread immediately through talking and also by mail which takes time to arrive).  For the synchronously updated SEIR and SEIS models used here, the inter-personal edges represent potentially disease-transmitting interactions that use agents' states at time $t-1$ to determine the states of other agents at time $t$, and so we will build a temporal web with cross-temporal inter-personal edges.\footnote{Other research in progress applies this technique to empirical data (e.g., interbank loan data) that includes a detailed description of building a temporal web with temporally flat edges representing all transactions within a time-span.  Though the analysis algorithms apply unchanged, there are difference in the constraints, caveats, and interpretations.}  

\begin{figure*}[ht]
\centering
\includegraphics[width=0.97\textwidth]{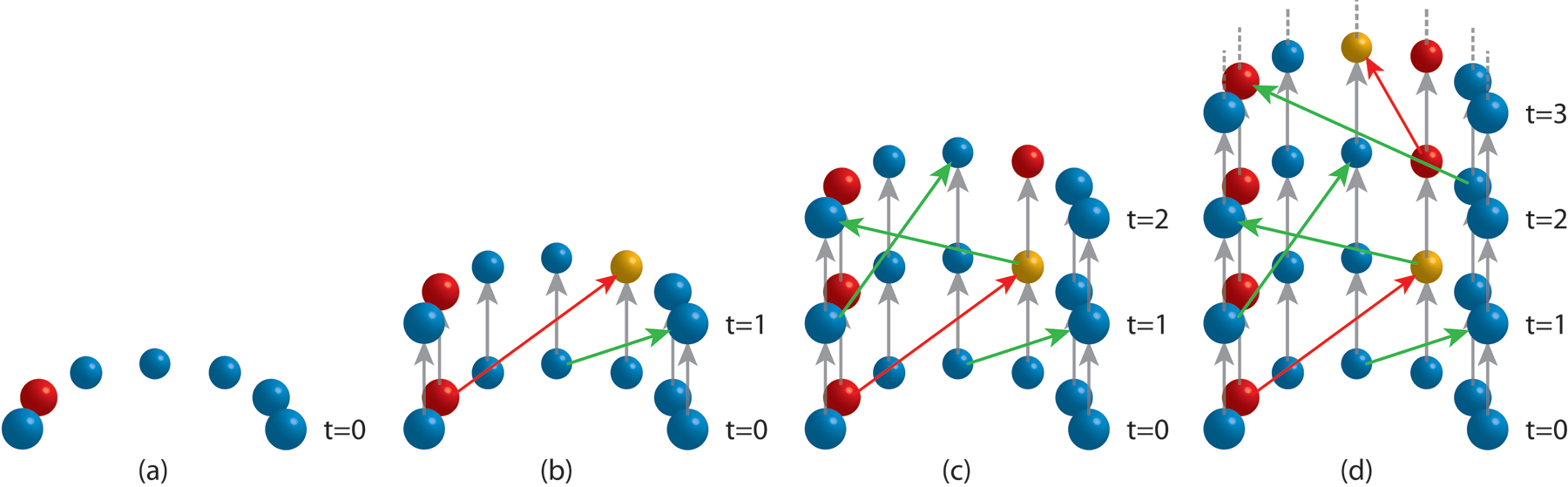}
\caption{A simplified example of building the temporal web from state-change and interaction data for the SEIR model. Note that the interaction edges are cross-temporal to capture simultaneous updating in the generated data.}
\label{building}
\end{figure*}

Figure \ref{building} demonstrates the building process with just seven nodes (referred to as nodes 1-7 from left to right). In panel  (a) the initial state of each agent is drawn for $t=0$.  According to the data used to build this temporal web, there were two interactions in the first period: from the infectious (red) agent 2 to agent 5, and from agent 4 to agent 7.  Disease-transmitting edges are colored red while safe interactions are colored green.  As a result of those transmissions agent 5 becomes exposed at $t=1$ and is represented as yellow in (b).  All other agents inherit their state from the previous period.  

This procedure is continued to period $t=2$ shown in panel (c).  The exposed agent 5 communicates with node 1 but it does not transmit the disease.  The data also reveals that between $t=1$ and $t=2$ agent 5 changed state from exposed (yellow) to infectious (red).  All other agents are again unchanged.

From $t=2$ to $t=3$ (d) we again have two interactions: from the now infectious agent 5 that exposes agent 4 (red arrow), and from the susceptible agent 6 to the infectious agent 2.  That this interaction \i{from a susceptible} agent \i{to an infectious} agent failed to transmit the disease is not an assumption of the temporal web, that is simply read in from the data (i.e. it is either true of the empirical data or a product of the model used to generate this data).  

The temporal web rendering process continues like this until all data points are represented.  Larger sets of agents, longer time periods, more interactions, and heterogeneities in agent numbers and interaction lengths across time add \i{complication} to this procedure (and time to the analysis), but the rendering process follows the same steps as this simplified example.

\section{Temporal Web Analysis}

Visualizing the crystallized dynamics of your model/data in a static temporal web structure may provide some immediate insights that cannot be realized from animations or equilibrium states, such as patterns in activity or transient effects. Although useful for conceptualizing a system, those benefits are limited and qualitative.  Because this way of generating a temporal web results in a single graph (rather than layers of graphs) existing measures from network and graph theory can be directly applied (with reinterpretation) to the temporally extended graph representation.  And because we chose an SEIR/SEIS model as our example, there are also available measures from epidemiology that can be derived from the temporal web, indicating useful interpretations for other systems captured in this way.  All that notwithstanding, a deeper understanding on the dynamics can be derived through a suite of analysis algorithms that have been adapted to harness the unique characteristics of this representation.

\subsection{Standard SEIR Measures}

Among the standard measures of disease virulence, prevalence, and morbidity the \textit{reproduction number} ($R_0$) is the dominant metric.  As succinctly stated in \citep{Rahmandad2008}: ``A central parameter in epidemic models is the basic reproduction number, $R_0$, the expected number of new cases each contagious individual generates before removal, assuming all others are susceptible.''  They report that a differential equation model using the same parameters we used for the ABM yields an $R_0$ value of 4.125, making this a highly infectious disease.


\textit{Cumulative cases} ($CC$) captures the number of individuals that ever become exposed or infected in the course of the disease.  As a measure of impact it captures the breadth of infection and approximates the total morbidity.  
The \textit{peak number} ($PN$) is the maximum number of agents sick at the same time.  Because in the SEIR model the recovery rate is near the rate of new infections, the number of agents that are exposed or infected (EI agents) hovers in a fairly narrow band compared to the cumulative number of cases.\footnote{Among all SEIR runs $\mu(PN)=24.54$ and $\sigma^2 = 15.10$ compared to an average of $95.90$ cumulative cases.  Out of the 1000 runs, 272 runs produce an infection with fewer than 30 cumulative cases and we refer to these runs as ``duds''.  Removing the duds from the analysis yields $\mu(PN)=32.35$ and $\sigma^2 = 9.24$ out of an average of 129.42 cumulative cases.}  For diseases in which $R$ stands for recovery, the peak number approximates the maximal impact the disease has on the functioning of society -- how many people are compromised on the worst day.
The time step at which the peak number is reached is the \textit{peak time} ($PT$) (if there are multiple time steps with a number of EI agents equaling the peak number, then it is the first of such periods).  

The temporal network approach facilitates another intuitive measure of disease morbidity that combines the cumulative number of cases and the length of their illness.  The \textit{temporal magnitude} is the proportion of nodes in the temporal web that are exposed or infected (EI nodes).  Because the nodes represent an agent at a time-step, the number of nodes ($N$) is equal to the number of agents ($A$) times the number of time-steps ($T$).\footnote{This holds as long as there is no agent entry or exit.  It is possible to incorporate birth/death processes with the appropriate modifications to node counting.}   Temporal magnitude for our application therefore equals $\frac{N_{EI}}{N} = \frac{N_{EI}}{A T}$.  For intuition purposes it may be helpful to note that when only normalized by the number of agents this quantity matches the area underneath a curve representing the percent of exposed and infected agents at each period.


Given the parameters and methods of the infection model used here, the length of the illness for each agent after first infection is determined by the probability of transitioning from E to I ($p(E\rightarrow I)=1/15$), and the probability of transitioning from I to R/S ($p(I\rightarrow R)=p(I\rightarrow S)=1/15$); these rates are constant and homogeneous across the population yielding an expected infection length of 30 time steps for each infection.\footnote{We observe a mean disease duration of 29.843 time steps in the SEIR model across all agents for 1000 runs.  In some runs, there are agents still in the E or I state when the runs terminate at 400 periods, thus truncating their disease duration.}  Because in this case the expected length of infection is homogeneous and known \i{a priori}, for the SEIR scenario the temporal magnitude is strongly correlated with the cumulative number of cases: the correlation coefficient is 0.9939.  

\begin{figure*}[h!t]
\centering
\includegraphics[width=\textwidth]{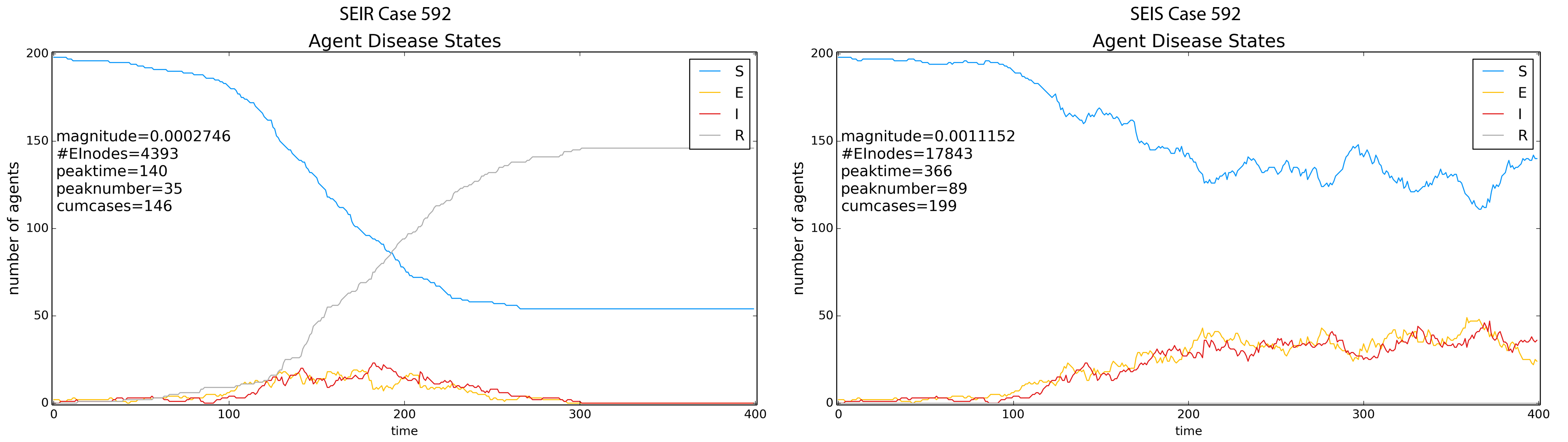}
\caption{The progression of the disease states as well as 5 measures of disease morbidity for run 592 of the SEIR model (a) and the same run of the SEIS model (b). The choice of this case is explained in section \ref{Results}.}
\label{measures}
\end{figure*}

However, for other applications in which either the cumulative cases or the average length are more complicated, temporal magnitude offers a simple measure of morbidity that is not reachable through combining other measures.  One only has to consider SIS, SEIS, and SEIRS variations within epidemiology to find such cases because recurrent illnesses do not increase the cumulative case measure.  In our SEIS experiments the correlation coefficient of the temporal magnitude with the cumulative number of cases drops to 0.9512.  This value is still quite high, however, this reveals that within the first 400 periods of the disease outbreak new agent infections are (unsurprisingly) the dominant driver of increased disease morbidity.  We found that it typically takes nearly the full 400 periods for everybody to become infected in the SEIS model given our interaction parameters.  Thus we can expect cumulative cases and morbidity to become increasingly uncoupled as time goes on.  Thus, in the contexts in which cumulative cases is appropriate, magnitude tracks it well; in the cases in which cumulative cases fails to capture morbidity due to reinfection, magnitude continues to measure morbidity.

Furthermore, when assessing the impact of quantitative agent states/properties across time (such as the debt of a bank, trade balance of nations, ...) the temporal magnitude measure generalizes to capture any attribute $x$ of the agents ($a_i$).  Summing a property's value across agents and across time will be informative only in certain contexts; for example, normalizing this by the number of nodes ($\frac{1}{A T}\sum_{t=1}^{T} \sum_{i=1}^{A} a_i(x_t) = \frac{1}{N}\sum_{n=1}^{N} x_n$) simply calculates the average property value across agents and time...a simple measure which has nonetheless demonstrated its usefulness time and time again.  For the SEIR and SEIS models, the average property is equivalent to normalized magnitude when exposed and infected agents have value 1 and susceptible and recovered agents have value 0.  This, we believe, is not a commonly presented measure of cumulative impact across time, yet it is an accurate representation of total morbidity and has clear applications to other systems (banks, nations, etc.).  It can be used to compare the total morbidity of different scenarios, in a way that improves over cumulative cases and peak number.  However, it cannot distinguish the different \i{timings} of disease instances, which we consider to be critical for understanding dynamical properties (and disease spread).

\subsection{Temporal Knockout Analysis}\label{TemporalKnockoutAnalysis}

In order to gauge the importance of each agent at each time for the spread of the disease, we propose a knockout sensitivity analysis inspired by knockout techniques used in genetic research \citep{Thorneycroft2001} and ecology \citep{Allesina2009}.  Temporal knockout (TKO) analysis determines how much a cross-time system property (such as total disease morbidity) changes when a particular agent is removed from that system at a particular time.  That is, for each agent at each time, remove it and run the system process holding everything else constant to identify what effect that agent has on the system \i{and when}.  The contingency of the system property to each agent-time's removal determines its \i{unique} causal influence on that property.

\begin{figure*}[h!t]
\centering
\includegraphics[width=.6\textwidth]{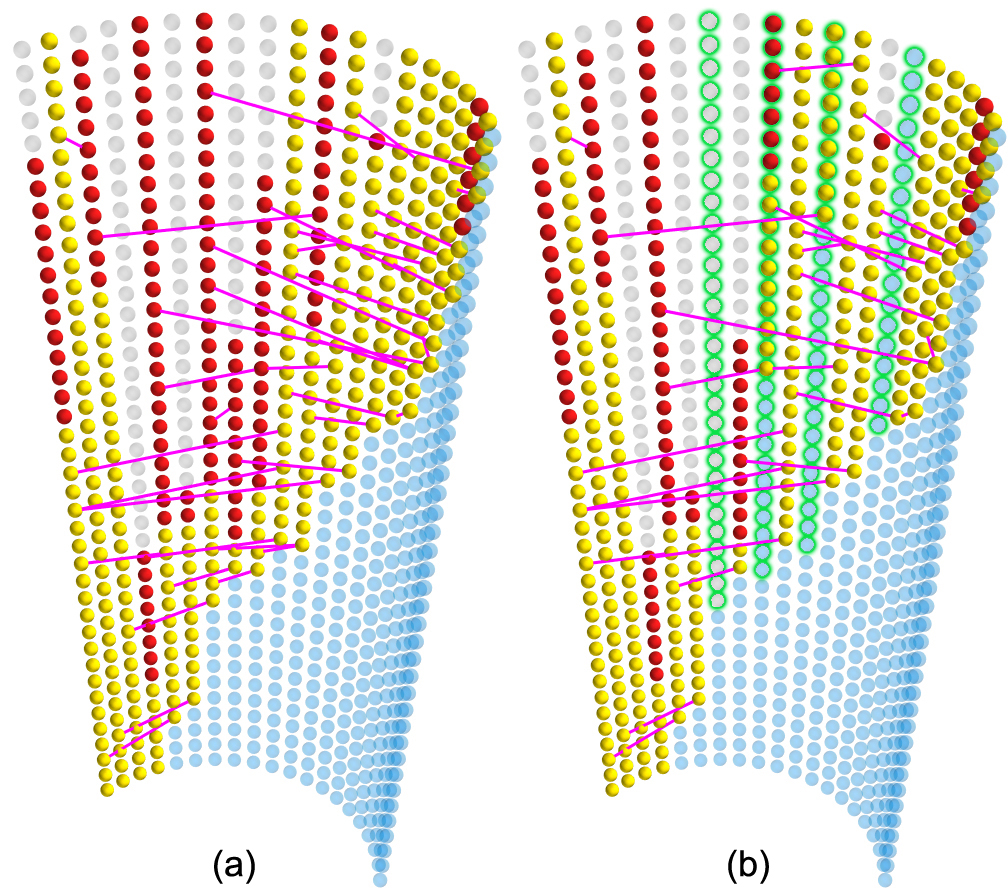}
\caption{The difference in dynamics between the base case (a) and the case in which agent 7 is removed at $t=10$ (b).  The normalized difference in the number of EI nodes (i.e. the delta temporal magnitude) is that node's temporal knockout sensitivity score.  Furthermore, there are variations in the exposed vs infectious profile (e.g. new positives generated vs infections averted) which can be picked up and differentiated by more fine-grained measures to produce multiple types of sensitivity.}
\label{TKOanalysis}
\end{figure*}

To achieve the ceteris paribus condition a preliminary step in performing a knockout analysis is to generate the \i{complete interaction dynamics} including who interacts with whom and which interactions would spread disease.  We call this the temporal network ``skeleton''.  For our models we generate an interaction structure based on the degree-depended interaction probabilities as well as store all the random numbers needed to govern the effects of interactions; i.e., whether each interaction would spread the disease if exposed or infected (which have different rates). We do this so that we can preserve the interaction structure and the contingent effects of each agent on the others while altering any of the infection parameters and disease states.  Then for each \i{node} in the system, we remove it (set it to the recovered state) while keeping the interactions the same. The effect this removal has on the temporal magnitude of the disease is our measure of the \i{sensitivity of the dynamics to that agent at that time} (see Figure \ref{TKOanalysis}).  We refer to this as the node's \i{temporal knockout score} or just \i{TKO} for short.  

The sensitivity of multiple system properties can be analyzed using temporal knockout, but we are focused only on total disease morbidity measured as temporal magnitude.  We believe that the TKO score is the best measure of an agent-time's influence on the system (at least for propagation models), and we therefore use it as the benchmark value against which all other sensitivity measures are compared.  Because the knockout analysis requires rerunning the simulation for all periods after the node in question for each node in the temporal network, it is computationally expensive: $O (A^3 T^2) = O(A N^2)$.  Considering this needs to be done for every run of a model, it is prohibitively time consuming for large numbers of agents and/or long periods of simulated time.  We therefore wish to find proxy measures that match the TKO rankings over nodes but with less computational time complexity.  Clearly any proxy measure for contingent total magnitude will need to span both time and the agents, and we are exploring the possibility that network measures across the temporal web will fill this role.

\subsection{Standard Network Theory Measures}

As described above, the directed edges connecting agents to their $t+1$ selves produces a temporally extended single graph (rather than a temporally layered series of graphs). In this application we have also used cross-temporal interaction edges to capture the simultaneous updating of the model, but even when the interactions are restricted to their time-slices the temporal web is a single network (though possibly a disconnected one) . Because the temporal web structure is also a network/graph, some common off-the-shelf measures from network and graph theory can be applied to it.  Naturally they must be reinterpreted to reflect the transtemporal meaning of the edges, and many common measures need to be adapted to work on directed graphs.  We will see that most existing measures fail to measure interesting properties of the dynamics under this formalism.  


It is also worth noting that while most measures on static graphs and temporally layered graph time-slices are agent-focused, our representation allows nodes that represent agent-times to be the focus of the analysis.  Although we also want to tie it back to an agent for identification and intervention purposes, and eventually to compare the dynamical properties to properties of the agents themselves, as an analytical technique the shift from agents to agent-time nodes fosters distinct calculations and insights.

\begin{figure}[h!t]
\centering
\begin{subfigure}[b]{0.45\textwidth} 
\includegraphics[width=\textwidth]{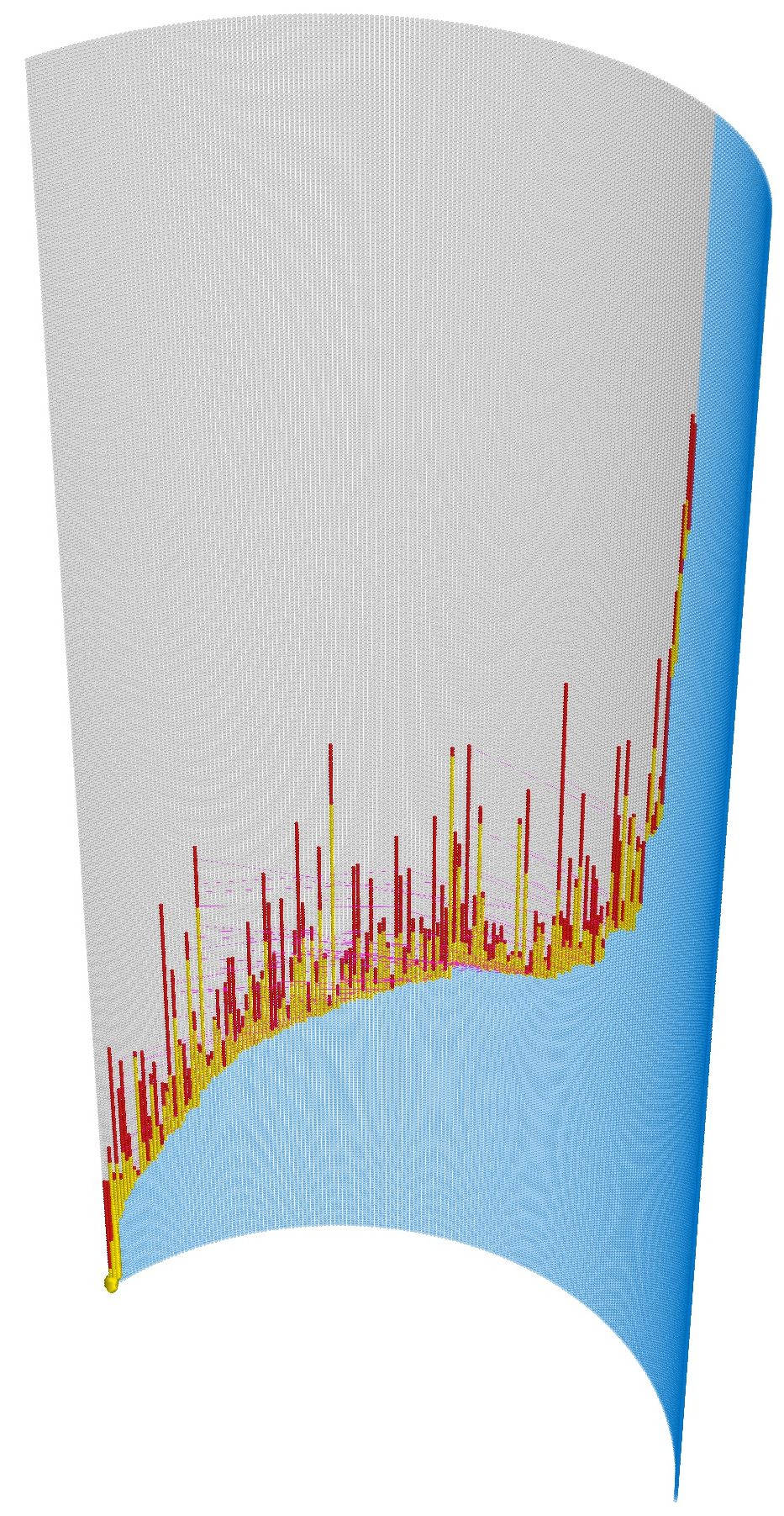}\caption{} 
\end{subfigure} \quad
\begin{subfigure}[b]{0.45\textwidth}
\includegraphics[width=\textwidth]{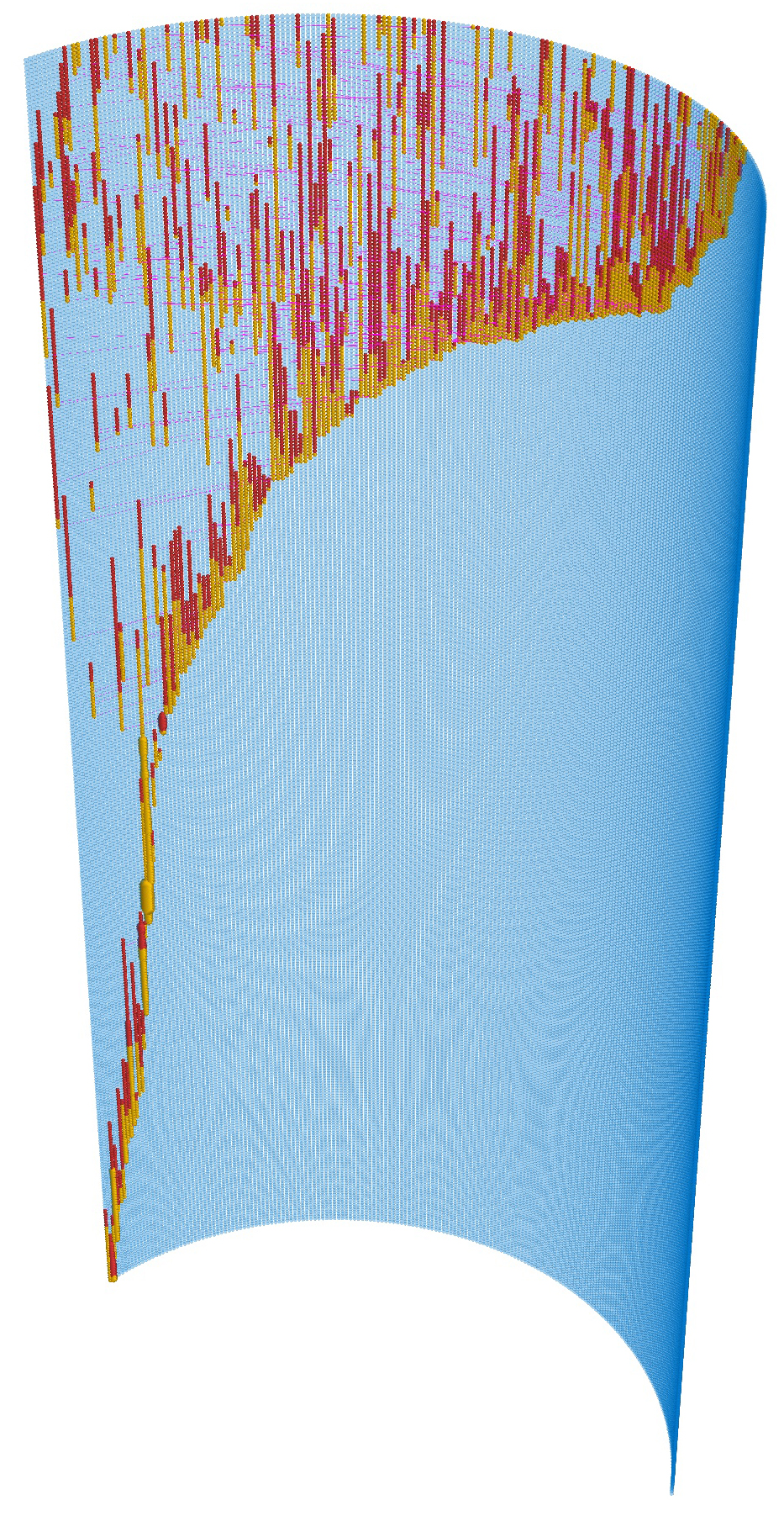}\caption{} 
\end{subfigure} \quad
\caption{The full temporal web of one run (case 848) of the SEIR model (a) and one run (case 338) of the SEIS model (b).  The yellow (E) and red (I) nodes together form the EI-subgraph representing the disease transmission in these scenarios.  Although difficult to see at this scale, the nodes are sized by their OCPaths/EI score (see section \ref{OCPdescription}) and these are cases in which that measure best matches the node-by-node TKO scores.}
\label{EISubgraphs}
\end{figure}

\subsubsection{Out-Component}\label{Out-Component}

The out-component of a node is the set of nodes that are reachable from that node by following directed edges. The definition is identical to the static (directed) graph case, but now it tracks influence across time through cross-temporal links.  In a temporal web the set of out-component nodes no longer represent specific agents, but rather all influenced agent-times of the focal agent.  In our analysis here we can restrict the range of our analysis to the \i{EI-subgraph} of the interaction structure.  The EI-subgraph consists of all nodes in state E or I together with the future-self connecting links between them and all the cross-temporal infecting links. Note that a cross-temporal infecting link can also connect to a node which already is in the E or I state.  The number of EI-nodes in a node's out-component captures the proportion of the disease magnitude that the node could have generated.  However, because there is typically a great deal of overlap in the out-components, this measure alone is not sufficient to capture unique, contingent contribution.  By incorporating various refinements we use the out-component as the base for many of our temporal web measures presented below.

\begin{figure*}[!ht]
\centering
\includegraphics[width=0.33\textwidth]{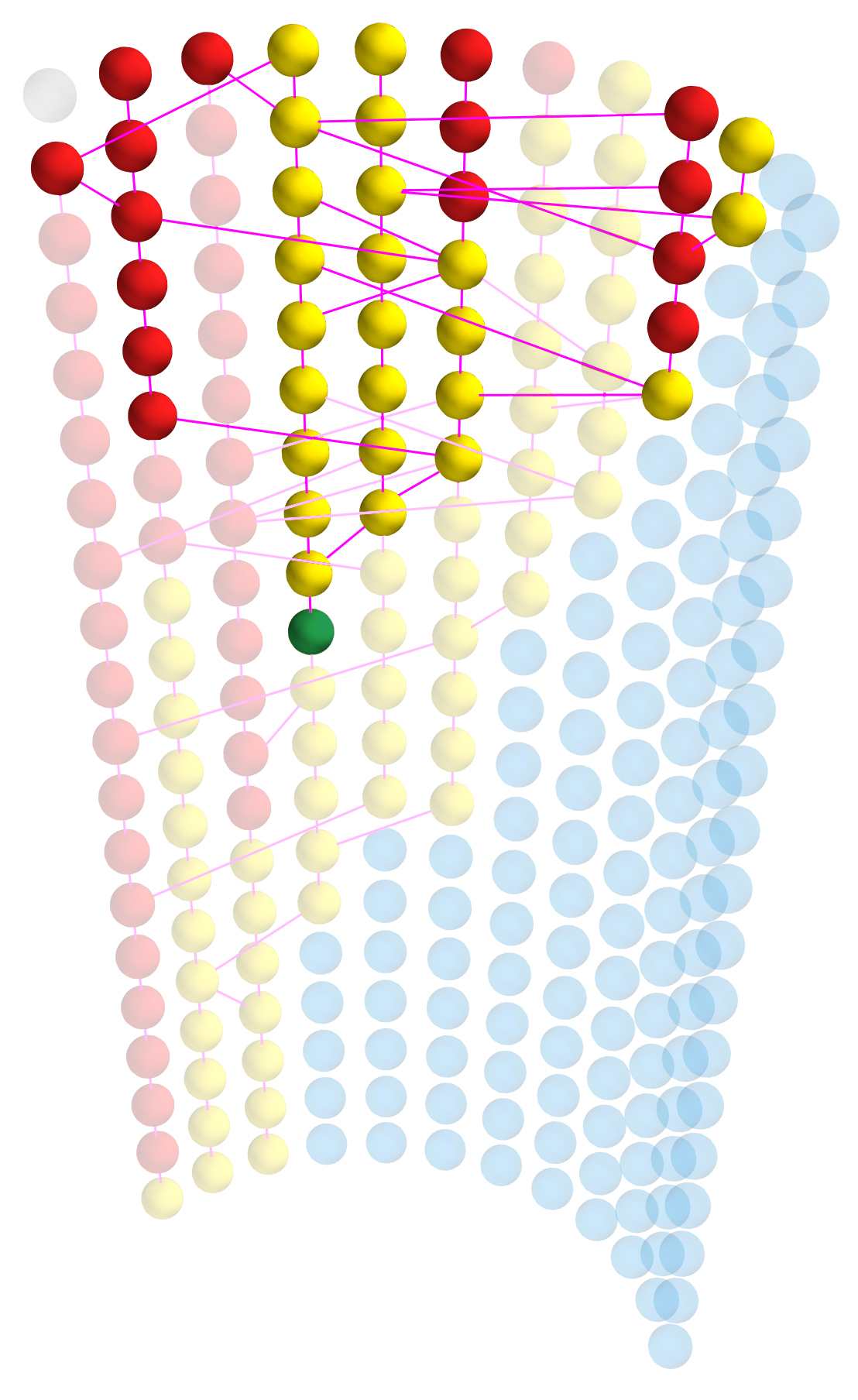}
\caption{The out-component of a node (green) in the isolated EI-node subgraph.  Note that in applications for which \i{not} all nodes are connected to their future selves (non-inheritance), the out-component may include a distinct set of individuals compared to temporally flattened graphs.  Furthermore, in the SEIS model agents enter and leave and reenter the EI subset over time; as such, one's future self at a particular period may or may not be in one's own out-component.}
\label{measures}
\end{figure*}

\subsubsection{Clustering Coefficient}

Standard measures of the \textit{clustering coefficient} report how frequently a node A is connected to a node C given that both are connected to a node B. The use of directed edges in this application already requires an alteration to the measure of the clustering coefficient, and cross-temporal interaction links makes true triangles impossible.    
Modifying this measure to track quadrangles of connection measures the frequency of interaction at time $t$ given an interaction at time $t-1$, a property which would be useful for many models (e.g. as a measure of the amount of preferential interaction). Doing so is equivalent to measuring the edge-overlap of adjacent time-slices \citep{Holme2012}, which is a notion of clustering, but not akin to the original ``friends of mine are also friends of each other'' purpose of the clustering coefficient measure.  For networks with interaction dynamics as sparse as our SEIR and SEIS models, the quadrangle measure is not (in practice) useful because the overlap of edges (in adjacent time-slices or otherwise) is frequently zero across the entire EI subgraph.

Capturing \i{temporal clustering} requires looking at larger inter-temporal patterns and determining how best to handle interaction directionality, time-induced directionality, and the span of time to consider.  Given a timespan, and whether interactions are intra- or cross-temporal, one can determine the number of possible interaction patterns consistent with a triangle in the flattened subgraph.  Each such pattern constitutes an inter-temporal network motif equivalent to the ``cluster'' measured by the clustering coefficient.  Figure \ref{motifs} in section \ref{MotifDetection} on our motif detection extension illustrates one such example.  A minimal cross-temporal example is: If A interacts with B and C in period $t$, do B and C interact with each other in $t+1$?  This has the same flavor of counting the number of triples that are triangles.  The case of A interacting with B in $t$, and then B interacting with C and C interacting with A in $t+1$ also creates a $3t$-spanning triangle, but it doesn't have the ``closing the triangle'' look and feel.  Most importantly, it will take considerably more work to determine if any of these triplet motifs provide useful information for temporal webs.


\subsubsection{Redundancy and Efficiency}

\textit{Redundancy} and \textit{efficiency} have technical definitions specific to network theory.  Redundancy measures how many of a chosen node's neighbors are connected to other neighbors of the chosen node -- for each node this is equivalent to one version of the clustering coefficient.  The ratio of the redundancy to the number of the chosen node's neighbors is the network's efficiency.  The lower the efficiency the greater the network density and number of redundant edges.  One can also interpret redundancy as a knockout measure: what proportion of the network's ``connections'' (not just edges) are lost when that node is removed.  As a measure of the robustness of network's connectivity it informs you of the number of nodes (or edges) that can be eliminated without compromising the propagation of the property. The weakness of these measures for temporal web applications is how the local node scores aggregate up to the global value. The kinds of reroutes that are implied by measuring redundancy on a flattened (or potential connection) graph are typically not possible on temporal webs.

However, the idea of measuring the robustness of propagation along paths can be measured in an alternative way for some temporal webs.  Imagine an SI or SEI model -- one in which nobody ever recovers (models of idea spread are often like this...no forgetting).  In such a model we can trace the propagation \i{along every path} from the first infected to all the nodes infected at the final time step.  Then for each agent-time node in the temporal web we determine what proportion of the paths are broken if that node is removed.  Low scores indicate redundant paths and low efficiency.  However, because the first nodes will always have extremely high scores, and the final nodes will have scores of zero, to measure a system's redundancy one will need to use the distribution of node scores to establish a redundancy/efficiency profile.  The null hypothesis would be that redundancy scales linearly with the time progression, and deviations from that indicate particular properties of the system's dynamics.  An approach similar to the Gini coefficient can then be used to compare redundancy across systems.  The use of such a path-knockout calculation has other uses beyond pure propagation sensitivity, but because we do not deploy them for the SEIR/SEIS analysis we leave deeper details to be described elsewhere.  As you will see in section \ref{OCPdescription} we do use a notion of path redundancy to augment the out-component measure, but it is not used to calculate a measure like the standard redundancy measure.

\subsubsection{Diameter}

The \textit{diameter} of a network is the longest geodesic path between all pairs of nodes; the minimal distance that ensures you can traverse the network.  For a temporal web the diameter's constraints depends on the specific construction.  If we put no restriction on the time-delay or time-span of interactions, then the diameter can be as small as 1 step when one of the initial nodes sends something to one of the final nodes.  If the agent interactions/relationships are within the time-slices (i.e., layered graphs), then the smallest possible diameter is always $T+1$ due to the essential directionality of transtemporal links.  If inheritance isn't assumed for the layered time slices then the longest finite length that the geodesic path could be is $T(A-1)$.  When inheritance is assumed the worst-case scenario is length $T+(A-1)$ because any agent that got revisited could just jump to its future self.  If the interaction links are cross-temporal links, as they are in our SEIR/SEIS models here, then the diameter is always $T$ (or infinity) because no path will use multiple edges per time step.

Determining the path length from each initially infected agent to every final agent across the EI nodes in an SI or SEI model reveals a combined measure of the time and the number of intermediaries necessary to reach the end state (or whichever time periods you choose).  By subtracting $T$ from the \textit{maximum geodesic time-spanning path} one can find just the number of intermediaries required for a time-order-respecting diameter of the flattened graph (obviously only useful on layered temporal webs).  In standard graphs, the diameter divided by the number of nodes is a useful, though basic comparison measure of overall network density/connectivity.  Whenever inheritance is assumed the normalized diameter reports the same feature while respecting the time-ordering of links.  When inheritance is not assumed, it reveals features of the interaction dynamics and connectivity that cannot be measured on flattened graphs.  And because a temporal web is build from interaction data (with no explicit network required), the diameter can be used to analyze differences in connectivity (e.g. small world effects) found in a dataset's or an ABM's de facto interactions.

\subsubsection{Closeness Centrality}

The standard \i{directed closeness centrality} of a node is calculated as the inverse of the sum of distances to all nodes in its out-component (or the harmonic mean to all nodes).  Because the graph is directed across time, it is never strongly connected, and therefore closeness centrality is intuitively of little use.  Suspending the directionality yields a measure of which node is most central to all other nodes in both receiving and sending influence, which may have some useful applications for information propagation.  The result is not a measure of influence because it traces backwards in time, but it can act as a measure of how ``in the thick of it'' an agent is at at time step.  Our measure of \i{nexus centrality} in section \ref{Nexus Centrality} is similar to this bi-directional closeness centrality. 

Note that when interactions are cross-temporal the length of any path connecting two nodes is the time difference between them, which simplifies the calculation algorithm.  In our SEIR/SEIS application the temporal edges dominate the temporal web's and the EI-subgraph's connectivity, so the later the period the closer the node is to all the nodes in its out-component.  In other applications, temporally adapted variations of closeness can be used to rank the importance of both agents and time periods for achieving a specific end-state by only measuring distances to the target time period(s).  For example, in the EI-subgraph of the SEIR model there are many dead-ends for the disease transmission. Tracking distance only along the infecting paths yields a measure such that greater closeness implies less long-term impact.  The ratio of closeness to the out-component size indicates the balance of multiple short paths versus a few long paths because the measures are cumulative across all nodes along the paths.  This and other versions are explored in future work.  

\subsubsection{Betweenness Centrality}

Ordinary \textit{betweenness centrality} is a measure of network structure that calculates for each node $i$ the number of geodesic paths between all pairs of nodes that include node $i$. Among other things, it indicates bottlenecks in the connectivity between communities of nodes \citep{Girvan2002}.  For the spread of a disease we are very much interested in bottlenecks of the sort that betweenness reveals in static connectivity graphs.  In a temporal web analysis, we would like to utilize bottleneck detection to uncover crucial junctures in the way interaction dynamics unfold.  

We applied the standard algorithm \citep{Brandes2001} and a directed-edge variation to each node in the EI-subgraph to determine each node's betweenness centrality in the system's infection dynamics.  Because the temporally directed paths only flow in one direction we believed that nodes near the end of the time span will typically have the most paths going through them as they accumulated across time.  However, what we found is that because betweenness only counts the \i{shortest} paths, the highest betweenness nodes occur in the busiest time steps (when the most inter-agent connections are found).  In our experiments nodes are usually not connected by multiple time-respecting paths in the EI-subgraph.  And when they are, the path length between nodes is always just the difference in time, so they are all the same length.  As a result, betweenness scores are the same or very similar across the nodes of an EI-subgraph in the current application, making them predictive of nothing.

Betweenness may be more useful for other applications, and it may be useful to make some alterations to its calculation for the temporal web domain.  For example, using layered interactions will produce longer and varied path lengths, giving betweenness something to track.  Going further, instead of calculating betweenness using the number of geodesic paths from all nodes to all nodes, we can consider a subsets of nodes; specifically, from one time period (e.g., $t=0$) to another period (e.g., $t=T$).  This \textit{timespanning betweenness centrality} may reveal how much \textit{and when} each element contributed to the transfer of the tracked property among the agents and across time.

We can refine our use of timespanning betweenness even further.  We can relax our constraint to the EI-subgraph to calculate the timespanning betweenness of the full graph to identify \i{potential} super-spreaders rather than actual spreaders.  This still differs from ``betweenness preference'' and other time-preserving betweenness calculations \citep{Pfitzner2013} by focusing on the propagation of properties across agent-times, but it becomes very close.  The set of nodes identified with the greatest timespanning betweenness value may correspond to what many refer to as a \textit{tipping point} or \textit{critical transition} in the dynamics \citep{Scheffer2012}, which should correspond to the TKO score in disease propagation.  That is, the identified nodes would mark an agent and time window in which removal (e.g. quarantine or previous inoculation) would be most effective.  Future work exploring the effect of various underlaying network structures will determine whether this or other variations of betweenness can capture bottlenecks in temporal webs.

\subsubsection{K-Core}

The k-core measure has been increasingly used as a measure of propagation sensitivity in networks.  For example, Karas and Schoors show that the k-core is most highly correlated with risk propagation in the time-flattened Russian interbank loan network \citep{Karas2012}.  It is also used in neural network analysis, idea (meme) spreading, metabolic networks, etc.  The sparse interaction among EI agents in our model makes it so that for most nodes the only edges are the temporal edges connecting agents across time.  However, those agents who undergo a short period of elevated interaction will have a large k-core robustness.  Although k-core does not perform well in tracking temporal knockout, it might be tracking some other dynamical property very well.  Future research will investigate what feature of interaction data, if any, the k-core tracks and where it may be uniquely useful.  We are also working on an adaptation of the algorithm to find the KT-core -- a time-integrating iterative k-shell removal technique to identify burst of activity in the temporal web. 

\subsubsection{Degree Centrality}

We capture the out-degree centrality of every node in the EI-subgraph with the idea that a node's individual contribution to disease spread may be best captured by its direct influence.  We also include the in-degree centrality to account for others' influence on it.  These two combined are measure of a node's throughput, which could indicate that a node is important for how the dynamics unfold.  Degree centrality performs well on temporally flattened graphs as a rough measure of influence, but on the temporally extended graph all our variations fails to capture the infection sensitivity.  The reason is that all nodes have the out-edge to their $t+1$ self and a few have one interactions, and a very few nodes have more than one interaction, and so every spreading agent will have the same out-degree.  Node degree fails because it only captures the immediate effects, and the kind of contingency that parallels temporal knockout requires incorporating downstream effects as well.  In section \ref{CumulativeDegree} we present extensions of degree centrality that further incorporate temporality through accumulating degree values.  Although node degree does not capture the property we are looking for, it is worth thinking further about what it might capture in other models and other constructions.

\subsubsection{Network Flow}

Network flow is less of a measure and more of a method to calculate measures.  There are many ways to construct and use a network flow algorithm, and many of our adapted measures presented below have an inspiration and calculation via network flow-type algorithms.  Determining critical agents with network flow is already common in networked epidemiological models \citep{Newman2002,Brockmann2013}, and it can also be used for community structure identification and robustness.  As with other measures the temporal directionality of our construct undercuts much of the power of network flow approaches.  With no cycles, and therefore no feedback, the standard measure is merely tracing the out-component, but it can be used in various alternative ways.  

For example, if we start flow from each node, let it follow all out-going edges, and add the flow at each node to account for all nodes that the flow passes through, then we can use the stored flow as a measure of message reception: how much of the system's flow passes through each node.  Looking at the full temporal web skeleton, we can use flow to determine the likelihood of nodes becoming infected for scenarios in which we don't know which agents will be initially infected.  Essentially this determines the number of time-ordered paths that a node is on as a measure of infectability.

But this is not a metric that is useful for disease propagation in which influence over the future is what is important.  For this application we can calculate the \i{time-reversed network flow} which scores previous nodes by their responsibility as a source of infection for each later node.  Or we can use it as another time-spanning technique that traces back to the actual or potential initial source for each unit of infection present at time $t=T$.  Such time-spanning measures are good for models without recovery, but fail to capture dead ends in SEIR or SEIS disease spread; and dead ends in the disease propagation still contribute to overall morbidity.  Hence network flow, normal and time reversed, is not a reliable proxy measure for temporal knockout here, but it is explored in depth in other work regarding the spread of technology and of bank risk leading up to a systemic collapse.

\subsection{Temporal Network Measures}

Graphs are tightly constrained representations that have been explored thoroughly over the past several decades. As a result there are few possible unexplored, simple measures of properties of networks.  However, variations in the structure of graphs (such as k-partite graphs, weighted multi-graphs, hypergraphs, and layered graphs) open up new avenues of analysis.  A temporal web as constructed here, however, is just a normal directed graph and therefore the measures that follow make use of familiar network property calculations.  Below we present a few measures devised to extract information about dynamical properties from the temporal web to demonstrate the sorts of mathematical gymnastics that make use of the temporal interpretation of the directionality.

\subsubsection{Cumulative Degree Measures}\label{CumulativeDegree}

Cumulative in- and out-degree is the amount of a system property that feeds into and flows out from an agent, accumulated across time.  For the SEIR model we can consider the property as 0 for the S and R states (not infected) or 1 for the E or I states (infected).\footnote{In a more nuanced application we could choose a value for the exposed states between 0 and 1 that captures the relative rate of infection spread for those nodes.  In general, it is not difficult to adapt these measures for real-valued node properties and/or edge weights.}  Summing in-degree for an agent going forward in time calculates how many times the agent has been infected up to each node's time-slice; we call this \i{uptake}.   The cumulative out-degree, or \i{discharge}, measures how many times the agent has already infected others up to that time slice.  The values of these measures at $t=T$ are the same as the respective flattened graph measures, and insofar as it tracks influence, the temporal web versions foster distinctions for when an agent's influence arises.   

Running backward in time, the reverse cumulative in-degree, or \i{popularity}, measure reveals how many times the agent will receive an interaction in the future.  The popularity measure could be useful for assessing the impact of vaccination schemes by informing us of both who and when a vaccination will protect an agent the most from likely infections.  Finally the time-reversed cumulative out-degree measures how many times the agent will spread the property (i.e., infect) to others in the future.  We can call this \i{gregariousness} which, like popularity, highlights the dispositional nature of measures over future behavior; these describe inputs and outputs if nothing intervenes.  

These four variations on the degree measure can be combined to capture a variety of dynamical properties.  Including the temporal aspect through cumulative sums allows these measures to go beyond immediate effects.  Intuitively one could combine the information regarding how many more times an agent will receive an infecting connection and how many agents it will contact in the future to determine the benefit of vaccinating that person at that time; e.g., by multiplying popularity and gregariousness.  However, due to the actual propagation patterns of diseases with our duration and infection parameters there are rarely more than a few interactions across time for any agent.  In our models an individual's impact on overall morbidity occurs through the indirect downstream infections, which are not revealed through degree measurements. For other uses of temporal webs, such as the interbank loan network, in which interactions themselves are important indicators of system health, these temporal web measures of cumulative degree may reveal themselves to be more useful.

\subsubsection{Out-Component Paths (OCP)}\label{OCPdescription}

\begin{figure*}[h!t]
\centering
\includegraphics[width=0.6\textwidth]{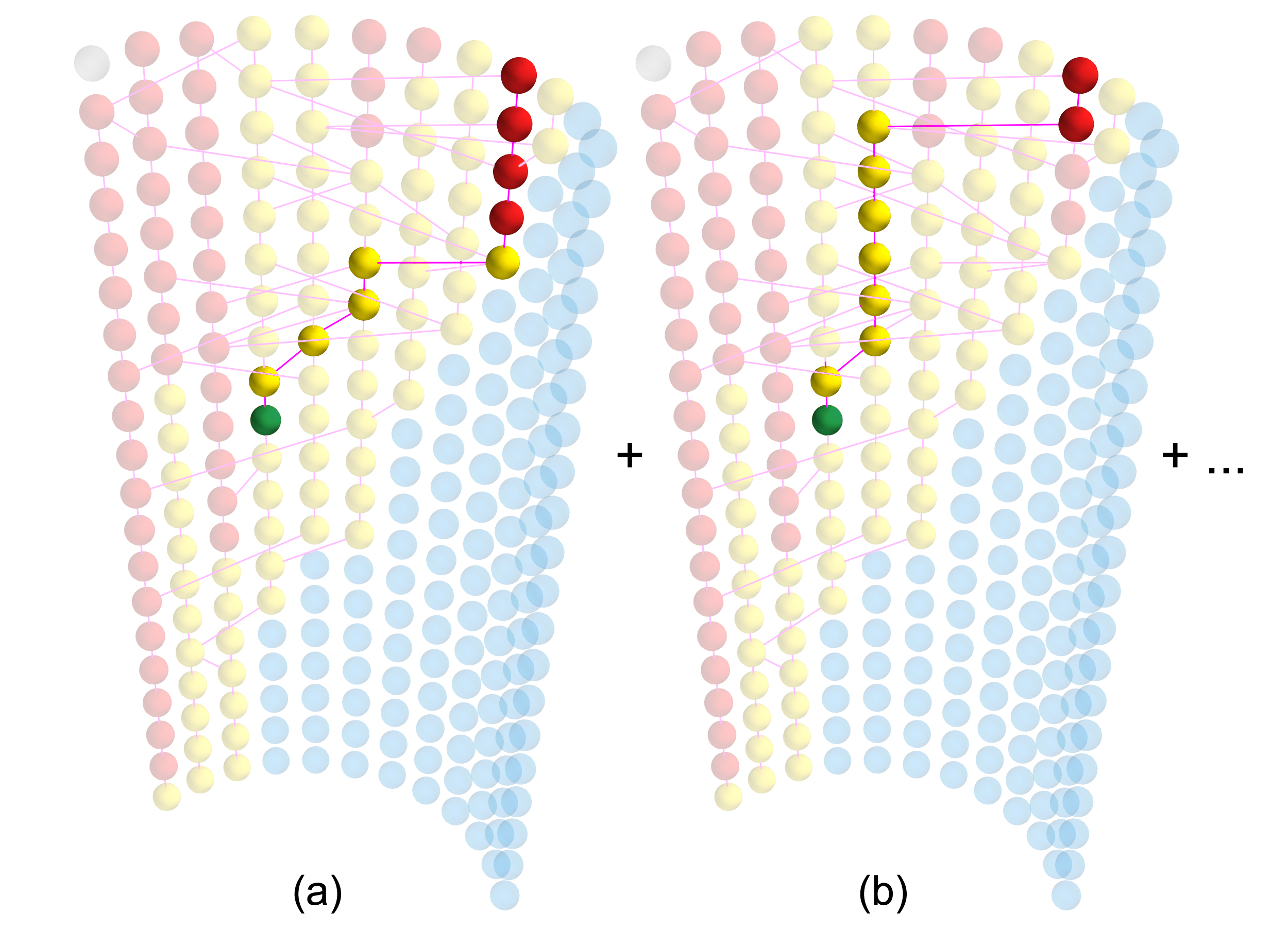}
\caption{To calculate the out-component paths score for the green node, we sum the lengths of all the paths from this node to each of the nodes in its out-component.}
\label{figure-OCP}
\end{figure*}

The standard out-component (described in \ref{Out-Component}) counts every node that is ``downstream'' of the focal node; i.e., every infected agent-time that this node may be responsible for (figure \ref{figure-OCP}). However, it turns out that this, by itself, fails to account for several other features that are important for capturing temporal knockout sensitivity. The feature of betweenness that we wish to capture in a dynamical measure is its ability to identify bottlenecks in the spread of the disease. The out-component actually under-represents the out-flow from a node because there are multiple paths from the focal node to many of the nodes downstream. In this case, we are not particularly interested in the shortest path connecting them -- any path will succeed in spreading the disease -- and each node along each path is a potential additional contribution to overall morbidity. Thus we sum the lengths of all paths from the focal node to each node in its out component.

By multiply counting nodes along each path we can better account for a node's potential, or contingent, impact on future morbidity. This allows a better match with the sensitivity analysis, but it is only a measure of output. To really capture the bottlenecks we want to measure not just high throughput, but high throughput with few alternative paths capable of maintaining that throughput. The problem with out-component paths is that there may be multiple agents with the same number of out-component paths (and many of the same paths) at a given time-slice. Although each one can be considered responsible for spreading the disease, none of them can be seen as crucial for spreading the disease. In order to match the sensitivities of the knockout analysis we further divide a node's out-component paths (OCP) score by the number of EI nodes that exist at its time-slice (OCP/EI).  This is still rather crude, and a technique that more carefully identified and adjusted for influence overlap may perform better, however we leave that analysis for future work because our results here show that it would be difficult to identify any measure that outperforms OCP/EI on these models.

Our algorithm for identifying all the paths for each node pair exploits the time directed nature of the temporal network structure.  We start with the nodes at the last time period (those with no out-edges).  Nodes at time $T$ trace back across each incoming edge and adds itself to each $T-1$ node's list of out-component nodes.  The $T-1$ nodes add themselves to their list of out-component nodes as they pass the list down to the $T-2$ nodes...and so on until $T=0$.  The time-reversed path tracing allows us to calculate all possible pairwise paths while crossing each edge of the temporal web exactly once.  For the case in which interactions are restricted to the immediate successor time period (such as our models here) the worst case computational time complexity for determining the out-component paths is $O(A^2 T) = O(A N)$.  Each node would need to add itself to the list of every node at the previous period and this would be required for every period.  

\subsubsection{Out-Component Paths Future In-Agents Weighted (OCPFIAw)}

As we will see in section \ref{Results}, the OCP/EI actually performs the best of the measures we developed, and better than any of the off-the-shelf network measures applied to the EI-subgraph. We also developed three other refinements that can be applied to the out-component or out-component paths that performed nearly as well in this application and may perform better in other applications.  These refinements also demonstrate the kinds of features one can look for in the dynamics captured by temporal webs, and hopefully inspire others to develop their own refinements.

\begin{figure*}[h!t]
\centering
\includegraphics[width=0.6\textwidth]{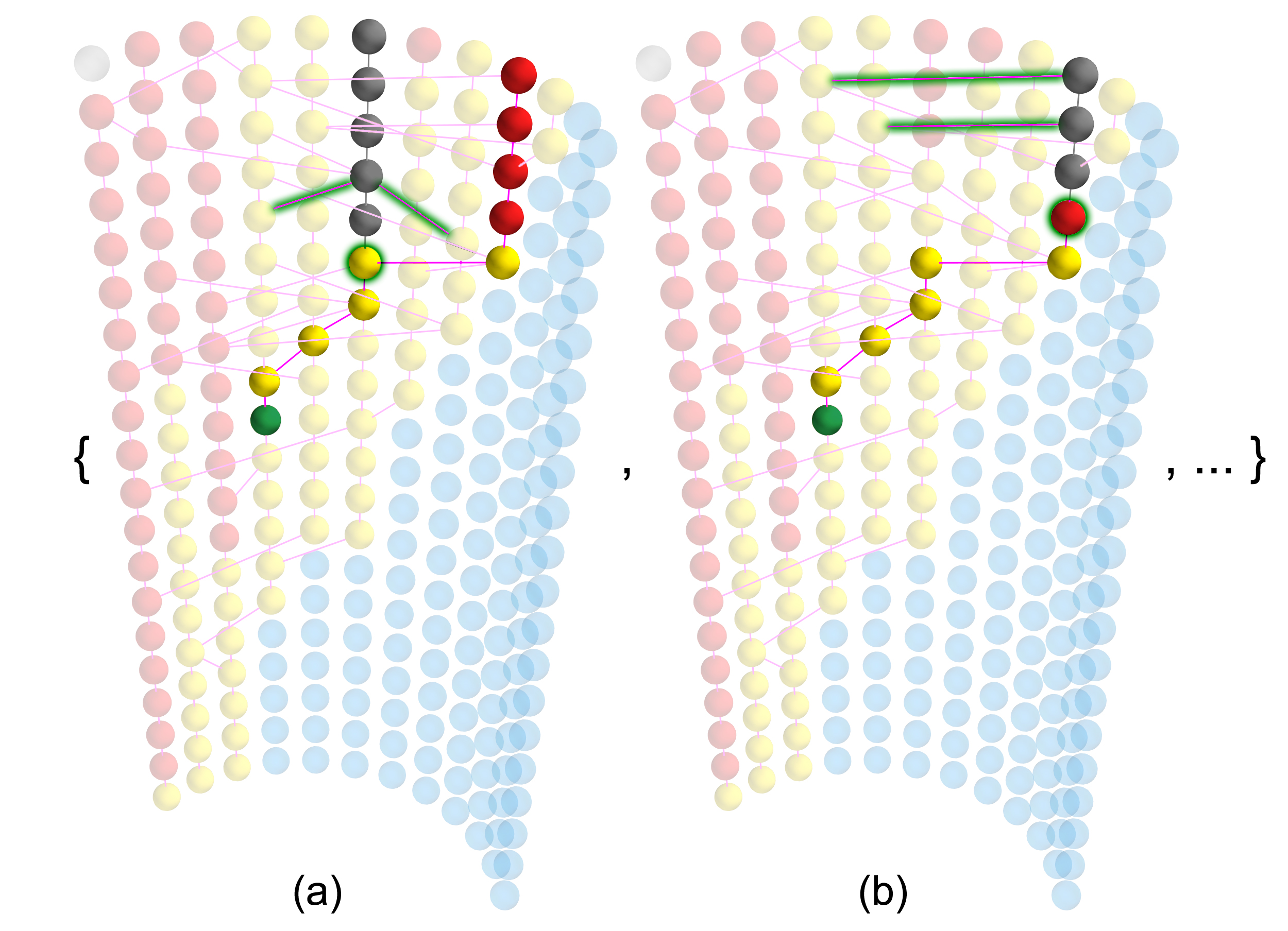}
\caption{Each node's contribution to the out-component path measure is weighted by the number of incoming edges (green glow) from other agents that occur in the future of this node's agent (gray nodes).  The greater the number of future in-edges from other EI nodes the less important this agent at this time is to the spread of the disease.}
\label{figure-OCPFIAw}
\end{figure*}

The first of these refinements weighs each EI node by the number of \i{unique agents} with an edge to that node \i{or any future self of that agent}. This measure connects the temporal web measures with a measure more similar to the temporally flattened ones. It utilizes unique agents and includes all future connections to the focal \i{agent} (rather than just that node). Our reasoning is that this node, if not infected by the path currently being evaluated, would still become infected via one of these future in-agents (figure \ref{figure-OCPFIAw}). Thus the contribution that this node adds to each of the paths using it is discounted by multiplying $1/(1 + $\#\i{FIA}); with \#\i{FIA} being the number of future infecting agents.

In a variant of this we also calculated a weighting based on \i{future in-nodes}, instead of agents (OCPFINw). The idea here is to capture the fact that those other agents have multiple infection paths running through them, and therefore counting it only once may fail to pick-up the disease contingencies properly. In practice, our SEIR-subgraph is sparse enough that the difference between in-agent and in-node weights is negligible. However, we mention it here both for completeness and because this adjustment (or a similar adjustment) may be precisely what is needed in other applications.

\subsubsection{Out-Component Paths In-Component Weighted (OCPICw)}

\begin{figure*}[h!t]
\centering
\includegraphics[width=0.6\textwidth]{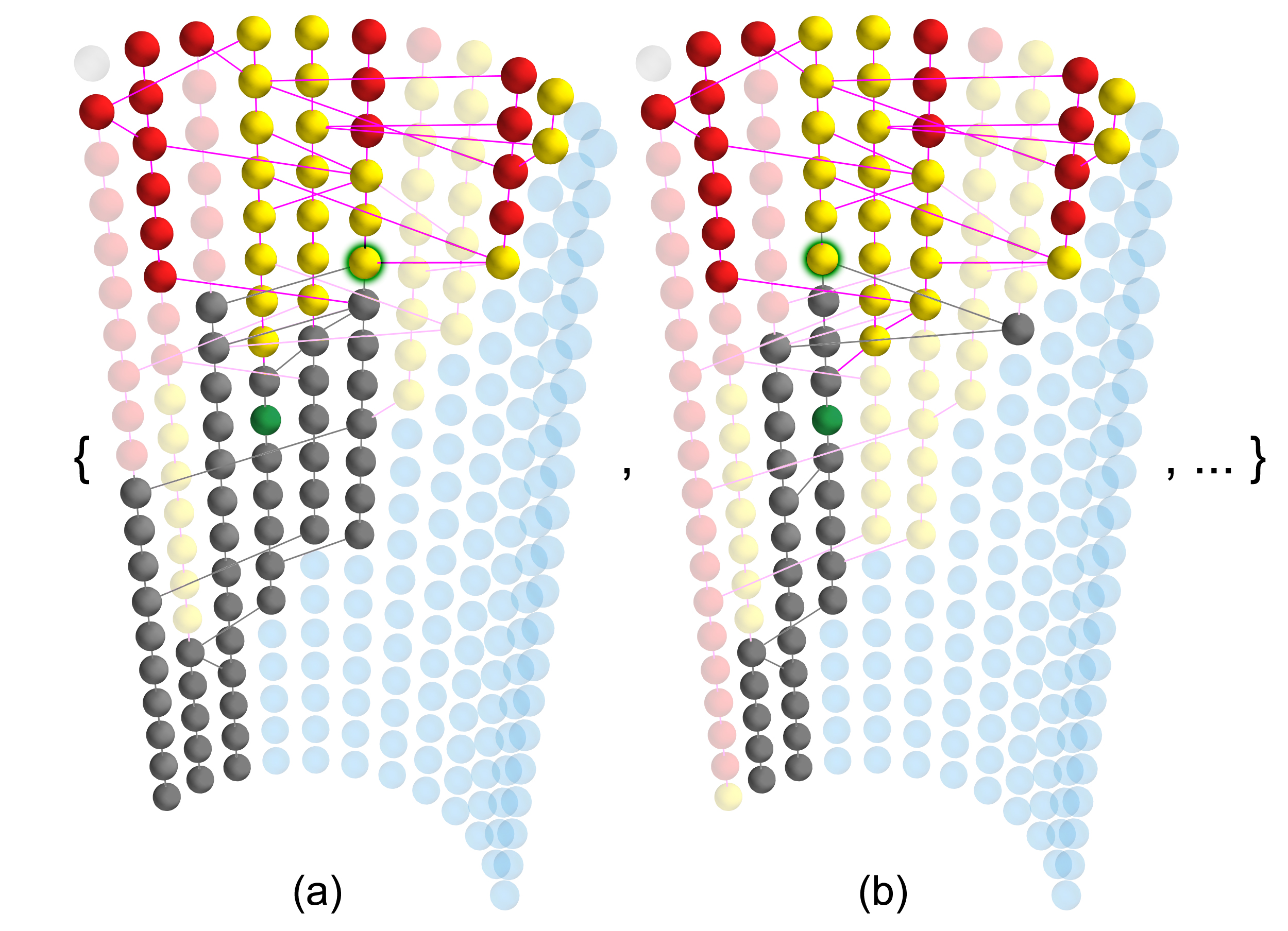}
\caption{Each node in the EI-subgraph is assigned a weight based on the size of its in-component with larger in-components decreasing the weight of the node.  The OCP measure is then calculated as the sum of these node weights across all paths connecting a focal node (green) to every node in its out-component.}
\label{figure-OCPICw}
\end{figure*}

Another approach to weighing the OCP nodes adjusts for the total level of influence received by that node.  For each node in the EI-subgraph we determine the size of its temporal in-component (all the \i{nodes} that have a path to it, $IC$). We then calculate each node's weight as $1 / ( 1 + |IC|)$.  Nodes with larger in-components add less weight to a node's temporal centrality because it receives influence from a larger base and therefore contributes less to importance of the focal node (figure \ref{figure-OCPICw}).  

Just as we have replaced the temporal out-component with the number of out-component paths, one could also calculate a modified version of this measure using in-component paths.  A further refinement weighs a node's contribution to a focal node's score by the proportion of in-paths to that node on which the focal node lays.  That is, if the focal node is on every path to the node, then it gets all its weight.  If several other nodes can reach it, then the focal node's importance is diminished.  Naturally there are further refinements to taking a node's accumulated receiving influence into account when assigning its contribution to other nodes' scores.

\subsubsection{Out-Component Paths Redundant Paths Weighted (OCPRPw)}

\begin{figure*}[h!t]
\centering
\includegraphics[width=0.6\textwidth]{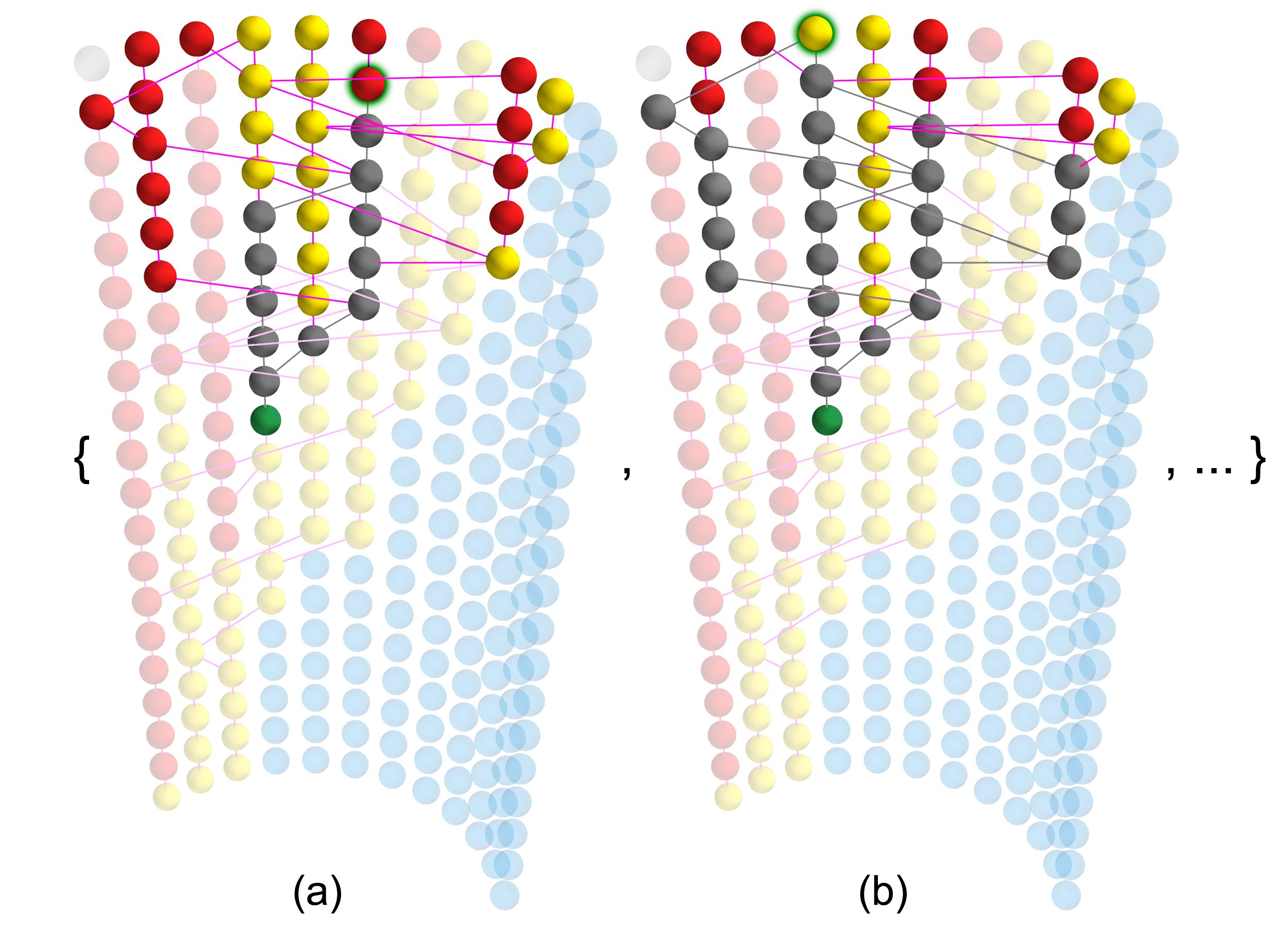}
\caption{For each node in the focal node's (green) out-component we determine all paths between them and the lengths of those paths.  This measure sums the number of node occurrences along all paths divided by the sum of the lengths of all paths each node is on.  This complicated measure produces a tradeoff that weighs short and/or unique paths more and long and/or redundant paths less.  }
\label{figure-OCPRPw}
\end{figure*}

Our most complicated, though not best performing, measure of temporal web centrality establishes a tradeoff between immediacy and redundancy.    We first determine all the paths from the focal node to each node in its temporal out-component. Now we use the lengths of those paths and assign a weight to each node based on the sum of the lengths of the paths on which it lies (figure \ref{figure-OCPRPw}).  Specifically, each node's contribution is the \i{number of paths it is on divided by the sum of the path lengths}.  In this way, nodes that lie on short paths count more than nodes on long paths.   If there are enough short paths then it can still equal the weight of a long path, but the idea is to discount nodes that contribute mostly to dead ends.  

This measure, and other variations similar to it, require a certain structure of the data to be useful.  For example, in this case there must be multiple paths of different lengths in the EI-subgraph in order for this measure to differentiate nodes by them.  What we observe for the SEIR model is a sparse interaction graph with few alternative paths as well as disease impact dominated by the few initially infected.  In our SEIS models the initial agents are still the most critical, although bottlenecks can exist.  These refinements will be more useful for models with denser interaction networks, more paths of different lengths, and longer running times, such as in our interbank loan network application.

\subsubsection{Nexus Centrality}\label{Nexus Centrality}

We also introduce a temporal measure related to closeness centrality and betweenness that is calculated from components and can be weighted by any of the adjustments just presented.  For each node in the temporal web (or in the appropriate subgraph) we determine all the paths in both the out-component and the in-component.  Then for each node we multiply (1) the sum of the lengths of all paths to all nodes in the focal node's out component (OCP from above) and (2) the sum of the lengths of all paths in the focal node's in-component (ICP).  This is the sum of the lengths of all paths running through the focal node.  As with the other centrality measures this can be utilized in various ways (e.g., using a subgraph of ``infected'' nodes or using a time-spanning measure from $t=0$ to $t=T$) as appropriate to the data and the information desired.  Because the in-component is not an important feature of disease spread we omit this measure in the results below, but preliminary results for identifying bottlenecks in ongoing processes (such as debt risk) are promising.

\section{Results}\label{Results}

We ran both the SEIR and SEIS agent-based model 1000 times and collected a full battery of measures: standard SEIR measures, standard network measures on the EI subnetwork, and our temporal web measures with various alternatives -- 79 measures total.  Recall that, as distinct from analyzing time-slices, we want to analyze the dynamics themselves by applying a measure to the whole temporal graph and identify dynamical properties.  Specifically we wish to identify those agents with the greatest impact on the disease spread and when they are key agents. We do this by comparing the scores for each node for each agent against its Temporal Knockout (TKO) score. As we explained in section \ref{TemporalKnockoutAnalysis}, the TKO score provides the best contingency sensitivity test because it simulates the same infection dynamics considering the elimination of each node and measures the resulting change in disease magnitude.

\begin{figure*}[!ht]
\centering
\includegraphics[width=0.85\textwidth]{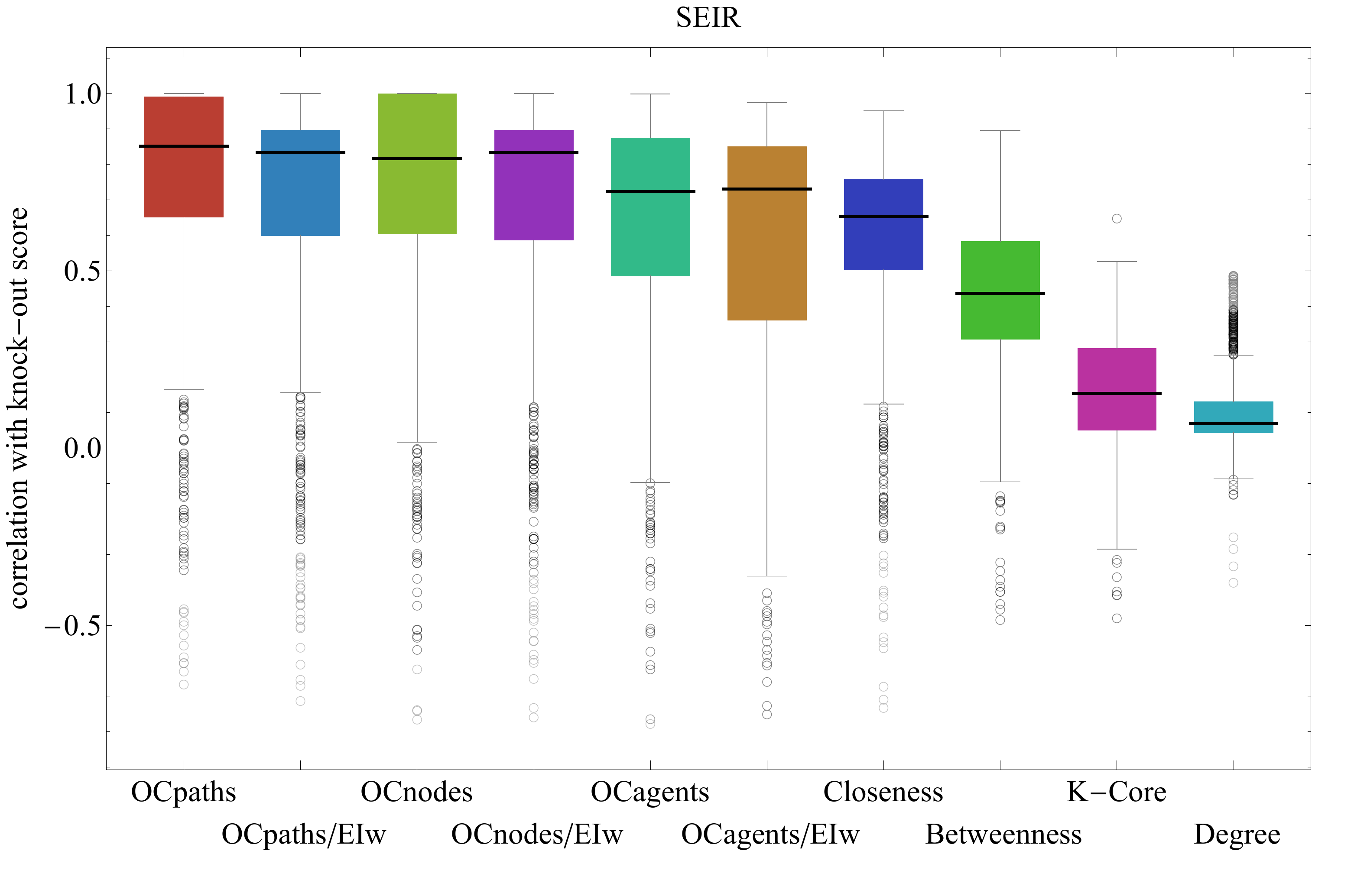}
\caption{The summary results from 1000 simulations of the SEIR model showing the node-by-node Pearson correlation between the temporal knockout score and several temporal web measures and off-the-shelf network measures.}
\label{SEIRResultsBoxPlot}
\end{figure*}

\begin{figure*}[h!t]
\centering
\includegraphics[width=1\textwidth]{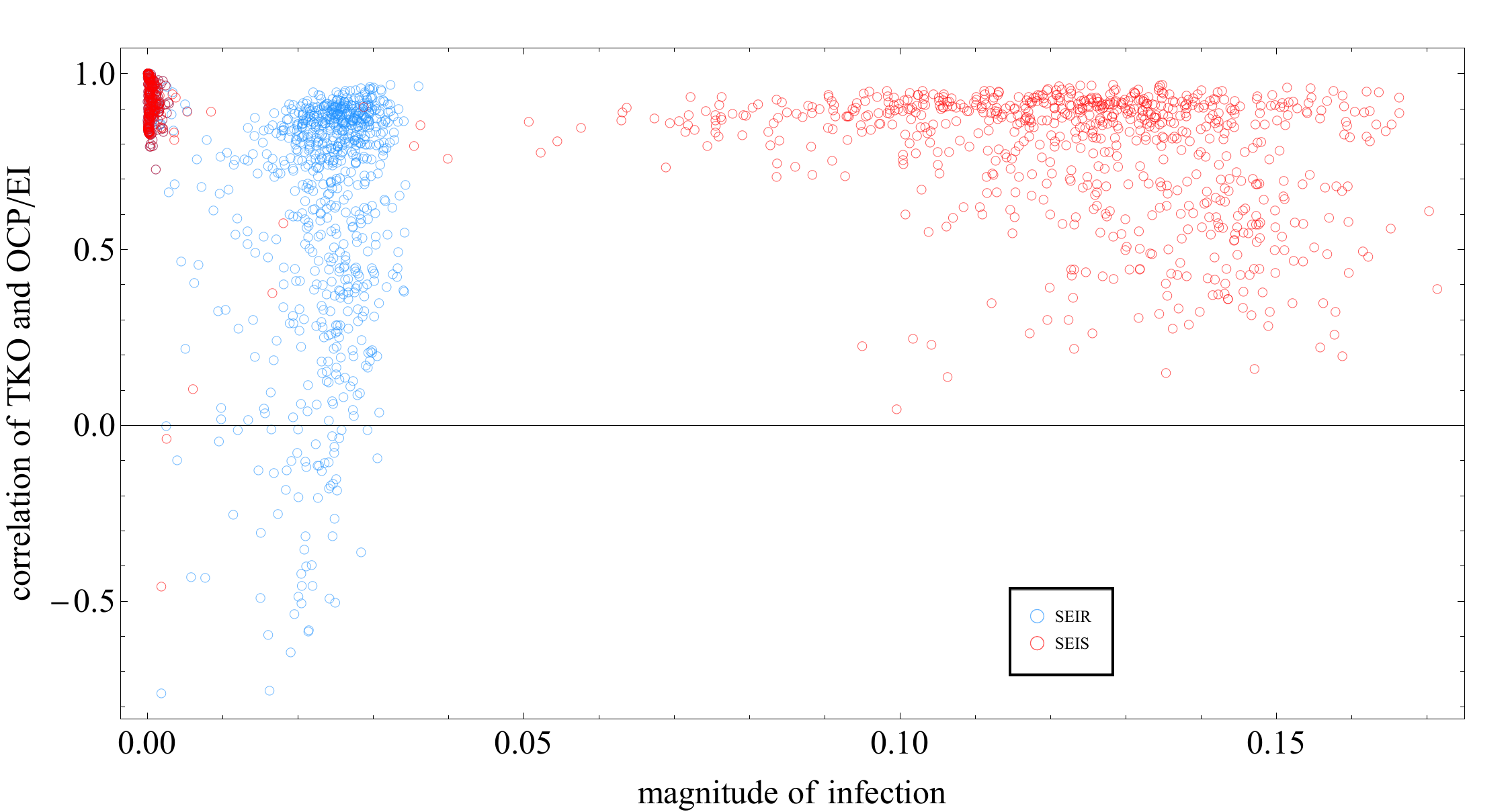}
\caption{This scatterplot shows the relationship between the morbidity of the disease as measured by the proportional size of the EI subgraph (temporal magnitude) and the correlation of OCP/EI and TKO.  When the diseases fail to spread we can easily identify the key, early individuals.  As the disease spreads more the performance becomes highly variable, and in some SEIR runs the OCP/EI measure is anti-correlated with TKO although it performs well overall in both scenarios.}
\label{OCPDIVEIMagnitudeResultsPlot}
\end{figure*}

Both SEIR and SEIS dynamics create a situation in which the initial agents typically have the largest impact on downstream infections.  This is not always the case, but when this is the case it is no surprise that out-component based measures will accurately capture that dependency. In our SEIR experiments 23.5\% of the runs infect fewer than 10 people (using the cumulative cases measures) and 27.0\% yield fewer than 30 individual infections (which we call ``duds''). Our SEIS experiments (which use the same random seeds)  similarly produce 23.2\% with fewer than 10 and 26.5\% with fewer than 30 individuals.  Our temporal web measures typically perform extremely well in these dud runs (clusters of points on the left of figure \ref{OCPDIVEIMagnitudeResultsPlot}) which may be upwardly biasing our performance results on more important scenarios. Specifically, we are interested in the the temporal web measures because they should be able to pick up properties of dynamics, but the properties we are interested in cannot exist in runs with so few infections.   We performed the measure comparison analysis with and without the dud runs, and the difference was too small to change the order of the measures' performance and therefore we present only the results with duds included.

Due to the large scale of our experiments and the number of variables we measured we provide only a summary of the overall results along with a detailed analysis of our best measure to provide the insight into how temporal web measures differ from traditional network measures and what they reveal.  Figure \ref{SEIRResultsBoxPlot} reports the correlations of ten measures across the 1000 runs of the SEIR model.  OCpaths and OCnodes perform nearly the same in both the standard and divided-by-EI-nodes-this-timeslice (/EIw) versions.  OCagents, which is equivalent to the standard out-component on the time-flattened network, performs only slightly worse.  
The off-the-shelf network measures all perform worse on average when applied to the temporal web, although closeness and betweenness do match TKO well on a number of runs.  For reasons explained in the network measure descriptions above, the standard K-core and degree centrality measures are poorly adapted to temporal web applications with cross-temporal interactions and that is revealed clearly in their poor performance here.  Closeness does as well as it does on the SEIR model because it assigns the nodes at the earliest time periods with the greatest closeness score; thus it is picking up a superficial property of SEIR dynamics on a time-directed graph rather than an important, general property of propagation dynamics.

\begin{figure*}[!ht]
\centering
\includegraphics[width=0.85\textwidth]{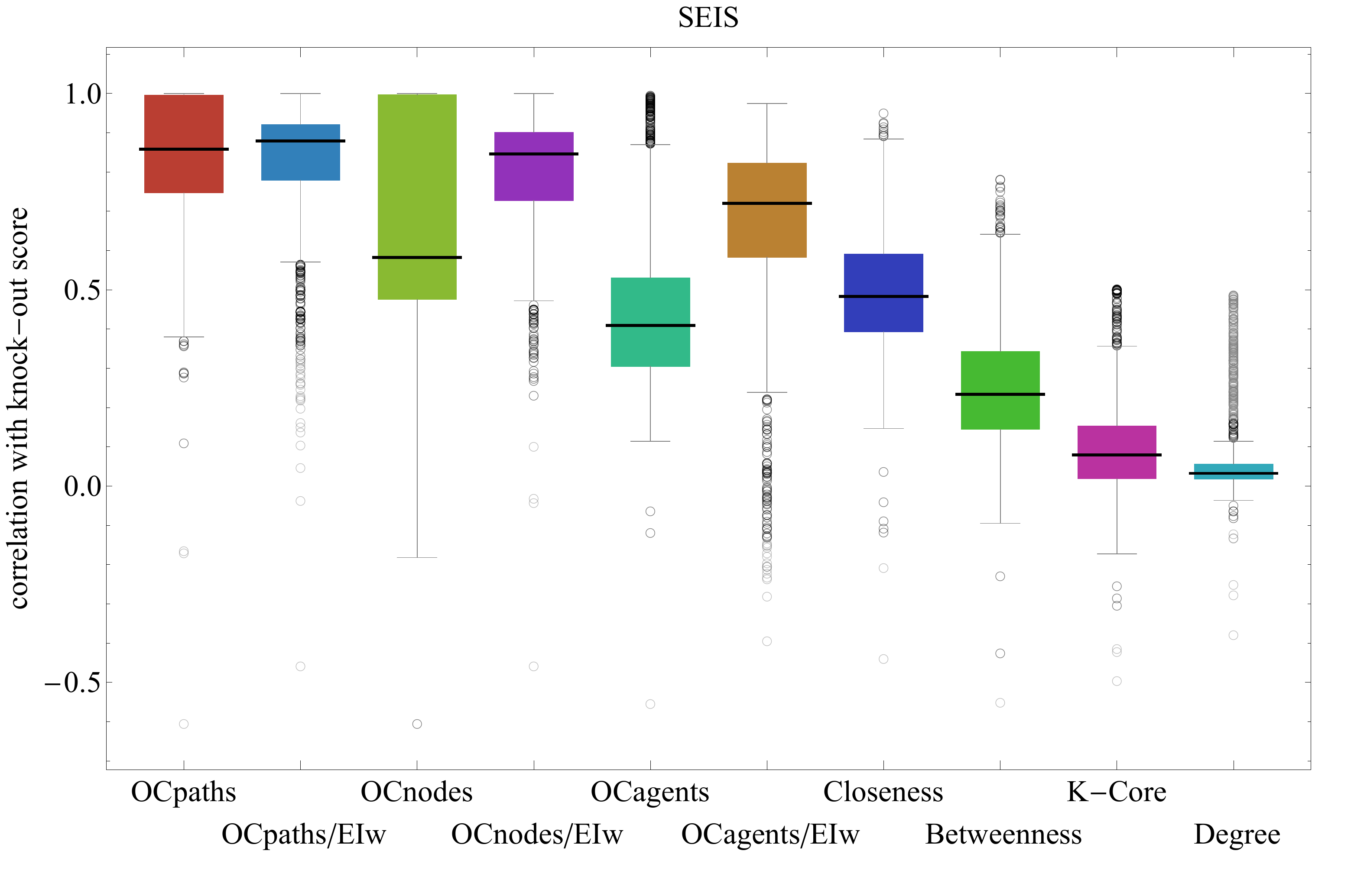}
\caption{The results from 1000 simulations of the SEIS model showing the node-by-node correlation between the temporal knockout score and several temporal web measures of dynamical properties.}
\label{SEISResultsBoxPlot}
\end{figure*}

As compared to SEIR models, the SEIS model has a much greater capacity for disease morbidity both in terms of the number of agent-times that can be in the EI state and the probability of infection through an interaction.  As time goes on in the SEIR model the pool of possible infectees decreases because more and more agents are immune, whereas an SEIS model on the same skeleton will use the same edge as a re-infection.  You can see the differences in magnitude in figure \ref{OCPDIVEIMagnitudeResultsPlot} in which the blue marks indicate SEIR results and red marks indicate SEIS results.  Greater magnitude mean greater contingency, and reinfection means that latter periods are more likely to be important.  Although in the growth stages of a disease, such as we modeled, the initial agents are still most likely to be the most important, the potential for a single infected person late in the process to generate a large magnitude is much greater.  Thus matching the TKO in an SEIS is more difficult.  

The results of our SEIS simulations are presented in figure \ref{SEISResultsBoxPlot}.  Here we notice that dividing by the number of EI nodes in a time slices drastically increases the performance of the temporal web measures in the SEIS model, although it had no consistent effect in the SEIR model.  Dividing by the number of EI nodes is intended to factor in the number of alternative paths of disease spread, and thus how bottleneck-like a node could be.  SEIR dynamics largely eliminate the potential for bottlenecks, but they are important for SEIS infection dynamics, and this difference is starkly revealed in the difference in performance of the ``$/EI$'' measures here.  

Also, as was expected, the OCagents measures (which are identical to the flattened network out-component) perform worse on the SEIS models because they cannot track increased morbidity from reinfection.  Closeness performs much worse than in the SEIR model precisely because early agents are less important in SEIS dynamics.  The other standard network measures perform consistently worse on the SEIS model as well.  OCpaths and OCpaths/EI both perform \i{better} on the SEIS model, although our various other weightings on the paths did not make any further improvements.  

The ability of OCpaths to track the greatest TKO agents so well can be understood through the features of the measure.  It combines the rough measure of total impact of the out-component with the ability to count the multiply infected and incorporate a measure of redundancy by counting the nodes along each path.  It can also be quickly calculated, making it a desirable proxy measure for the complete knock-out analysis.  By examining the performance of the measure in more detail we may be able to identify those features of the dynamics that effect the performance the most, and thus further improve our ability to capture the dynamical properties we are interested in.

\begin{figure*}[h!t]
\centering
\includegraphics[width=1\textwidth]{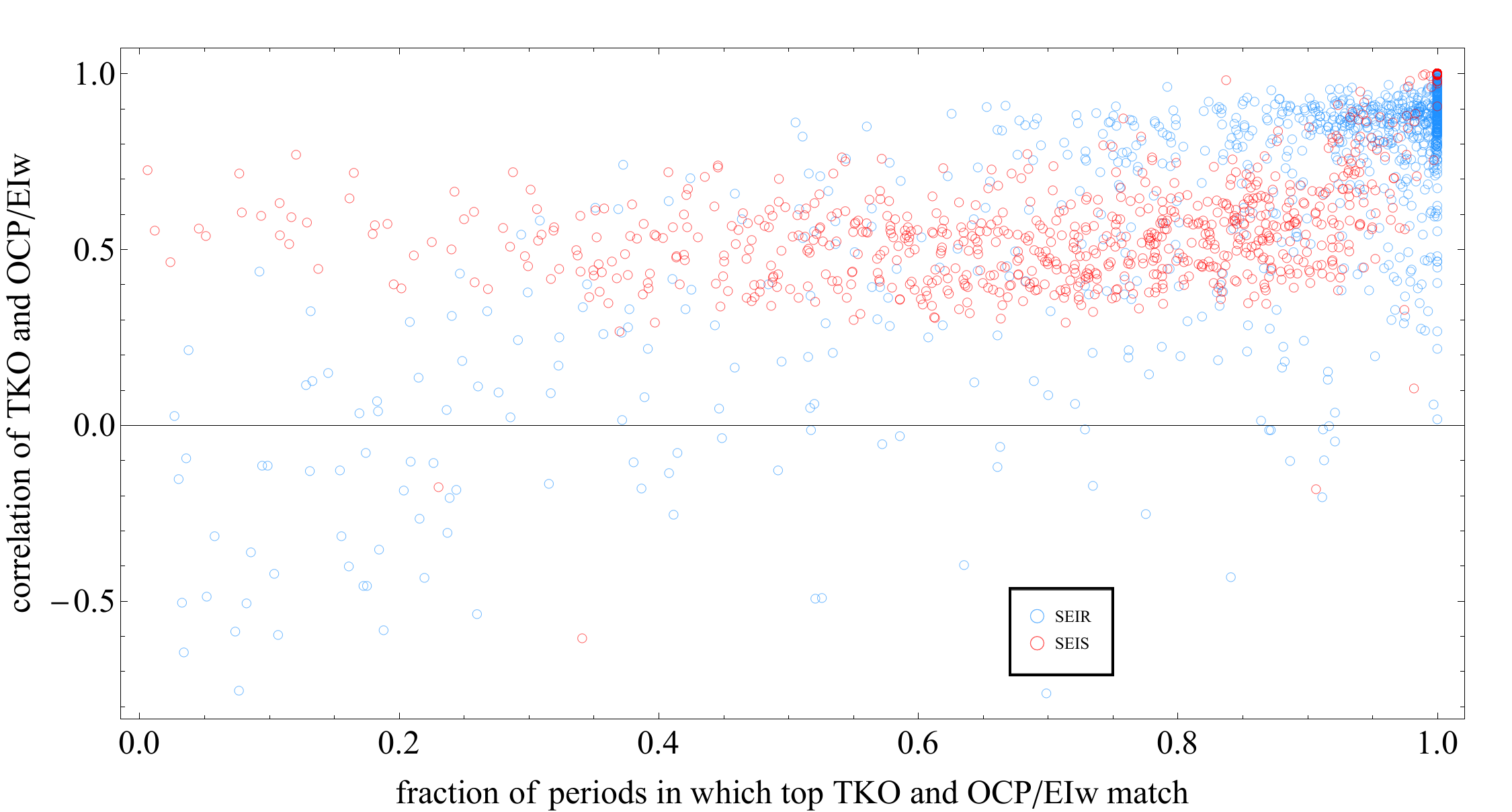}
\caption{This scatterplot shows the relationship between identifying the highest TKO agent (x-axis) and the node-by-node correlation with the TKO value (y-axis).  The variation in correlation cannot be sufficiently explained by matching just the most sensitive nodes, thus indicating an overall good fit.}
\label{OCPDIVEIResultsPlot}
\end{figure*}

Figure \ref{OCPDIVEIMagnitudeResultsPlot} shows the relationship of the magnitude of infection and the performance of OCP/EI measured in terms of Pearson correlation with TKO.  Figure \ref{OCPDIVEIResultsPlot} instead compares the correlation with the fraction of periods in which the OCP/EI measure accurately identified the highest TKO individual.  This latter performance measure is meant to reflect the usefulness of the temporal web measures for selecting the best target for intervention policies.  If multiple agents could be vaccinated, quarantined, or otherwise removed from the system, an overall high correlation between OCP/EI and disease sensitivity indicates that our measure is generally a reliable proxy, and better in SEIS scenarios than SEIR scenarios.  Comparing the two figures shows us that high correlation almost always parallels matching the top agent each period, and overall lower correlation is not strongly tied to the magnitude of the infections.  Specifically, once one removes the duds, a linear regression reveals no significant positive or negative relationship between magnitude and correlation whereas the matching fraction and the TKO correlation are strongly positively correlated for both the SEIR and SEIS data.  

\begin{figure*}[h!t]
\centering
\includegraphics[width=\textwidth]{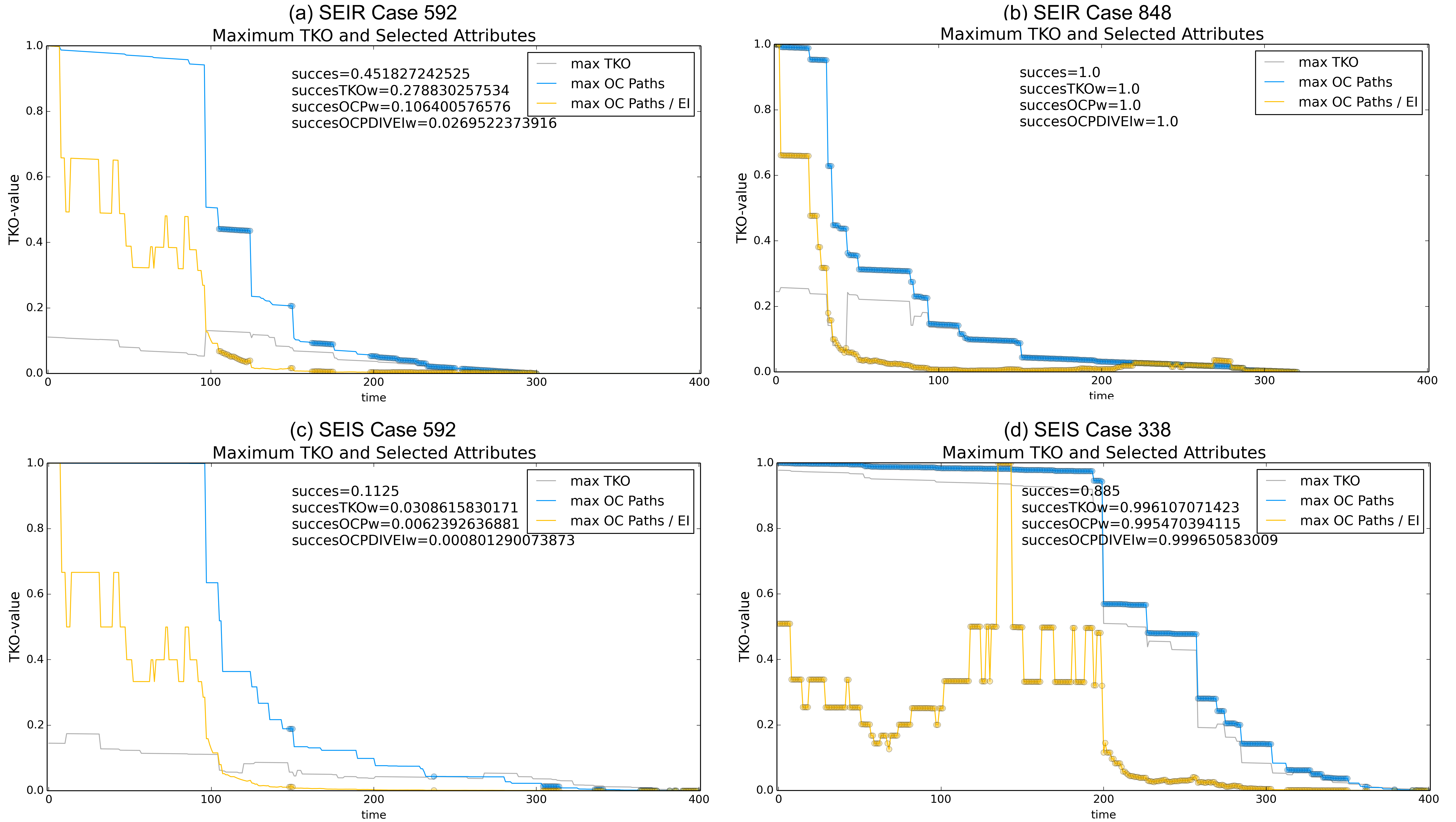}
\caption{This plot shows the success of OCP/EI, our overall best measure, to identify the highest TKO agent at each time step as well as demonstrate some variations in the TKO patterns.  The embedded scores indicate (from top to bottom) the raw proportion of OCP correct matches, the OCP match proportionally weighted by the TKO value, the OCP proportion weighted by the OCP value, and the OCP/EI weighted by the OCPDIVEI value.  Time steps with disc marks indicate a match between OCP and max-TKO.  }
\label{TKOMatchPlots}
\end{figure*}

For our simulations the overall fit, rather than only matching the highest nodes, was important to determine because we found that the gap between the highest and second highest score per timeslice was often quite large.  The property of carrying the greatest TKO and OCP value appears to move from agent to agent across interactions as time progresses, only rarely are there multiple agents with high values at the same time step.  Thus it seemed possible that matching only this highest agent could explain most of the correlation, our investigation into the fraction of max-TKO matching shows that this is not the case. Furthermore we did not find and consistent patterns  in the relationship between max-TKO levels (samples shown in figure \ref{TKOMatchPlots}) and overall measure performance.

The four cases in figure \ref{TKOMatchPlots} show the best-case and worst-case scenarios for OCP and OCP/EI in both the SEIR and SEIS models.  Actually there are many cases in which the OCP and OCPDIVEI measures match perfectly, so I have presented here the cases with the greatest temporal magnitude for each one.  Case 592 is conspicuously the worst-case for both models.  Recall that both models use the same skeleton; i.e.; the same random seed and the same sets of interactions.  The \i{only} difference between these two scenarios for case 592 is whether agents become susceptible or recovered.  And for both models this case produces nodes with negative TKO values.  When a node has a negative TKO score this means removing that agent at that time \i{increases} the total disease spread.  This can happen when the interaction structure is arranged in a way such that if some particular agents are infected early on they quickly recover without spreading the disease, but the same agent infected later will stay infected and infect many others.  Identifying them is tricky because it is not the focal agent being knocked out, but rather one of the agents that the knocked-out agent infects.  That is, a node will have a negative TKO if at least one of the agents it infects (directly or indirectly) would produce greater disease morbidity if it were instead infected at a later time.  And not just greater than that too-soon-infected agent created, the increase must be greater than all the magnitude removed by knocking out the focal node.  The reality of this counterintuitive possibility is partly responsible for the low performance of case 592, and it also points to interesting feature of SEIR/SEIS dynamics still in need of accurate measure.  


Our results show that despite the less-than-perfect match overall and high variation across some runs, our out-component paths and related measures do outperform standard network centrality measures on the EI-subgraph.  Recall that this is not a comparison with these network measures applied to the temporally flattened graph equivalent of the dynamics (that is coming in future work).  We applied these off-the-shelf measures to the time-directed EI-subgraph of the dynamics.  The directed edges of this network, as well as the sparseness of interactions across time, account for the poor performance of the standard measures in this context.  Nascent techniques for time-layered networks cannot be applied here because there are no layers in version of a temporal web with cross-time interactions.

Path-finding and network flow approaches to measuring dynamical properties seem to offer the best results, however the field of temporal networks is still young and better measures may be found as it matures.  What is probably more important to keep in mind is that these measures performed better \i{at the task of tracking disease impact across random interactions}.  Changing the task will shift which dynamical properties are relevant, and that is going to change which measure is best at capturing the desired property.  It is for this reason that we presented not just our best measure and the SEIR/SEIS results, but also other varieties of measures and our thinking behind them.  Researchers interested in other properties of other models are invited to use, adapt, improve, and reinterpret the measures provided here.  The SEIR/SEIS results are indicative of the benefits of temporal web analyses, gaining additional insight into such well-tread territory.

\section{Extensions and Modifications}

As already mentioned, this paper presents the temporal web measures with interpretations appropriate for the SEIR and SEIS models.  Applying the temporal web technique to other models, data, and problems makes other measures more relevant.  There are, however two, important additions to make to the methodology that can be explored on the SEIR and SEIS structure.  The first extension is to refine the interaction protocols from random mixing to specific mixing patterns.  Second, we are interested in gaining more insight into patterns of dynamics by explicitly identifying recurring substructures in the temporal web; i.e., detecting intertemporal motifs.  Although the application to simple models is useful for refining and demonstrating the technique, we plan to apply the technique to available temporal network datasets both to evaluate our approach in comparison to others and to provide useful information regarding the behavioral contingencies and dynamical properties.

\subsection{Underlying Network Structure}

The current analysis utilizes homogeneous agents in order to to focus on the ability of tracing the greatest impact agent and time using only the revealed interactions; i.e., a scenario in which no other system information could potentially inform us of the outcome contingencies.  In addition to the fully connected base network presented here (which produces uniform random interaction probabilities) we are also interested in exploring the effects of heterogeneous interaction patterns.  This can be done using a nonuniform distribution of interaction probabilities or using structured underlying interaction topologies from the literature; for example by using random, small world, scale-free, and ring lattice networks as the potential interaction conduits.  This future work will allow us to compare the typical static network properties of agents in the underlying potential interaction structure (and/or the temporally flattened, observed network structure) to their TKO score and our temporal web measures.  Such an analysis will reveal whether an agent's temporally flattened network properties (such as standard degree or betweenness centrality) really operate as reliable proxies for their dynamical properties and system influence.  

In principle we could run the static network measures on the social network implied by the temporally flattened graph of the observed connections in these models to test how well standard network measures perform in comparison.  Actually we did do that, but with these interaction parameters the temporally flattened graph is nearly fully connected in every run (average edge density is 97\%).  Standard network analysis is useless here, but exploiting the temporal element fosters further, though different, analysis.  Where standard network theory has shined is on more structured social interactions.  By (1) restricting interaction along pre-determined pathways, (2) exploring the effects of several different initial conditions on each skeleton, and (3) comparing the actual disease impacts to the impact expected from the standard flat-graph measures we will be able to test the accuracy (and hence usefulness) of these standard measures for propagation dynamics.  We may discover, for example, that different network characteristics are important at different stages of the disease spread, and thus different interventions are recommended.  This research may also help us identify new measures for flat-graphs that perform better in identifying key agents in social networks.



\subsection{Dynamical Motif Detection}\label{MotifDetection}

Network motifs are subgraphs shared across network structures and/or recurring within the same network.  Being subgraphs, they are formed from a subset of a network's nodes and their connections, sometimes augmented with particular properties.  Structural motifs have been examined in the context of gene regulatory, metabolic, neurological, and many other biological and artificial networks \citep{Kashtan2004,Milo2002,Sporns2004}.  By identifying patterns of importance in the structure of a network, and highlighting all the occurrences of those patterns, motif detection can aid in establishing the building blocks of a larger structure, find key relationships in the structural or functional fabric, fill in missing links, and other descriptive tasks.  The idea of using dynamical motifs to capture interesting features of network dynamics appears sporadically in the literature (e.g. \citep{Zhigulin2004}), but the approach here is novel to the temporal web construction.  

\begin{figure}[!ht]
\centering
\includegraphics[width=.3\textwidth]{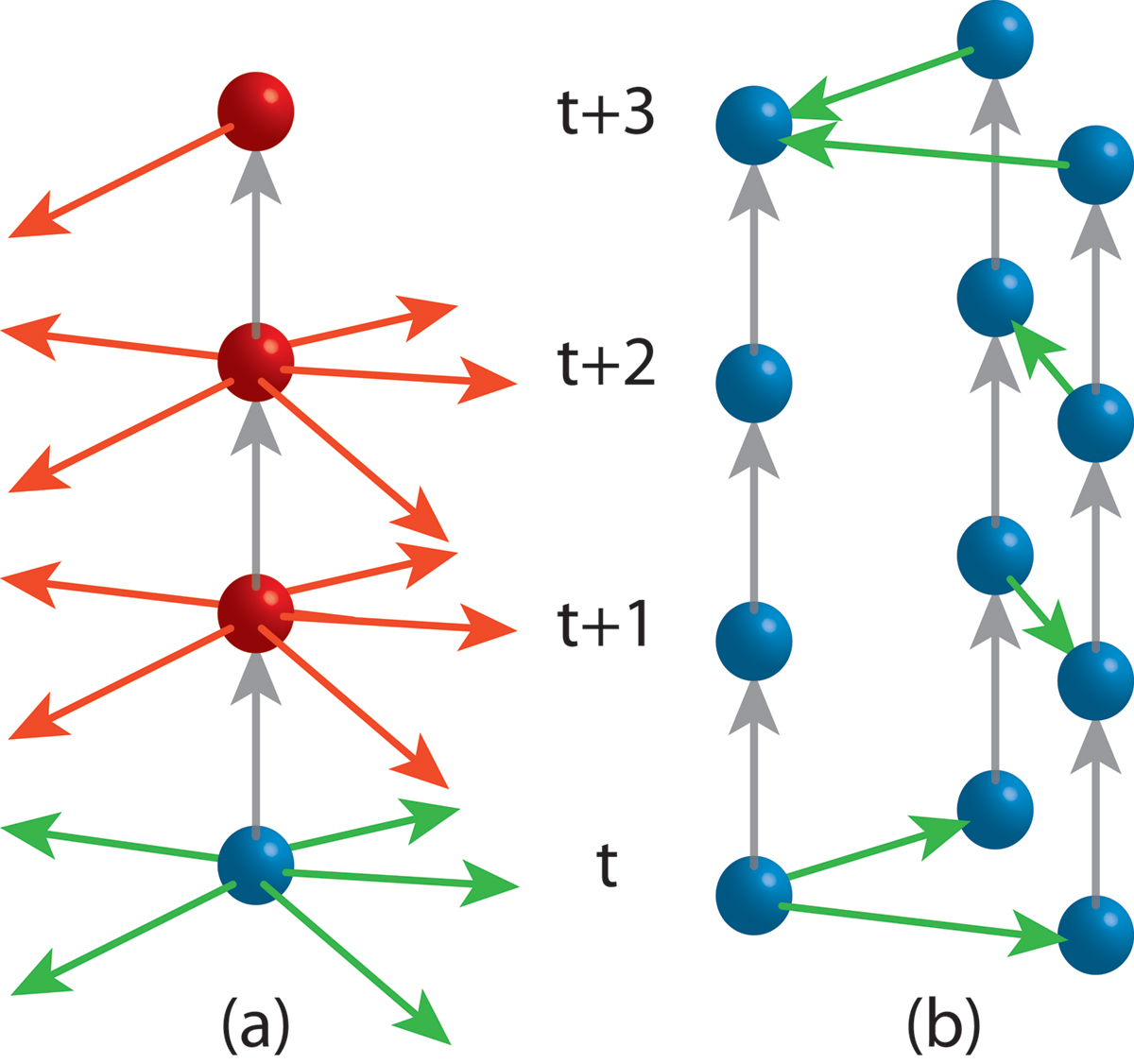} 
\caption{Dynamical motifs using intra-temporal relationships representing (a) a reduction in interaction from 5 to 1 occurring two periods after an agent becomes infectious, and (b) a pattern capturing questioning, discussion, and answering (among other behaviors depending on context).  Note that the flattened version of (b) is the complete directed non-reflexive graph of three nodes, but this is only one of 4096 possible structures that match the complete flattened graph using a 4-period time span.  Motifs are shown here using intra-temporal interactions for ease of presentation, but they can just as easily be defined for cross-temporal interactions.}
\label{motifs}
\end{figure}

For the temporal web, the motif itself is static and only requires traditional motif detection algorithms to capture and match.  The time-directed structure actually simplifies the search algorithm.  When the motifs include intertemporal or crosstemporal interaction edges, the motif represents a pattern in the dynamics of the system's state changes and interactions.  For example, if we wanted to test the hypothesis that people reduce their interaction rate a couple periods after becoming infectious, then this can be represented as a class of motifs in which the degree of nodes is higher before and immediately after becoming infectious than it is two periods later (an example of such a motif is shown in figure \ref{motifs} (a)).  Thus hypotheses about individuals' contingent behaviors, or the contingent interactions among several individuals, can be tested for using this technique.  

Dynamical motifs may involve multiple elements over time to capture patterns of interaction and changing relationships.  If there exists a repeated interaction structure -- such as repayment of debt, answering questions, reciprocity, passing along information, etc. -- then that will appear as a particular series of connections across time.  Collecting the behavioral motifs of a system and comparing them across parameter changes (and even distinct domains) can be used to discern categories of system behavior.  By using \i{motif schemata} (structures including ``wildcard'' elements similar to those used in genetic algorithms and/or ordinal timing of interaction events) we can eventually generalize recurring themes in system dynamics and establish a catalog of intertemporal interaction patterns.  Thus the shared representation as a temporal web facilitates identifying the cross-domain patterns in the dynamics of systems, an important development for the theory of complex systems.

Aside from identifying building blocks and coherent behavior, dynamical motifs can be used to make predictions over future system behavior.  Given a library of known motifs, incomplete patterns can be matched over (for example) the first $\delta t - 1$ periods of the motif.  Then one can determine the probability that the final period will complete that motif based on either historical completion frequency or based on all possible next steps (i.e., assuming nothing about likely connection patterns).  Such a predictive tool goes beyond exploiting cycles and chains in the network dynamics to basing expectations on a large repertoire of previously observed or theoretically important system behaviors.  And because the temporal web can be built from any collection of time-series interaction or relationship data its usefulness extends beyond systems currently modeled as network structures.




\subsection{Applications to Empirical Data}

We are already applying this technique to empirical inter-bank loan data from Russia \citep{Karas2010} to detect systemic risk and better understand how it accumulates. 
We are interested in uncovering specific properties of the interaction system: how do banks mitigate risk, absorb risk, and push risk off to other banks?  Under what conditions do bank failures lead to cascades, and are their patterns of behavior that can reliable halt such a cascade.  In collaboration with other groups we are also pursuing applications to activity on a brain connectome mapping, Twitter retweet data, online and real life social networks, idea transmission, as well as additional simulation models on a variety of topics.  Insights into the behavioral patterns of these systems will come from understanding the dynamics (not the equilibria) and that is why we need improved measures of the dynamical properties.


\section{Conclusion}

We introduced several measures of dynamical properties for identifying the time and agent sensitivity of disease spread in typical SEIR and SEIS models.  This simple epidemiological demonstration provides a base from which to explore the dynamical properties of more sophisticated models.  The nature of propagation often imbues early infections with the greatest importance for the spreading of a disease in both SEIR and SEIS situations, but not always.  Our temporal knock-out technique is capable of determining the exact actual contribution of each agent being infected at each time by measuring the change in social morbidity resulting from removing that agent at that time.  The out-component and path-related measures perform well overall in identifying the key agents and key times for the disease propagation -- and provide quantitative measures of their importance.  


For other models the spread of the property of interest will depend much less on the initial agents and more on bottlenecks through the process.  This is true of SEIS dynamics after the disease has reached a certain threshold of infection and becomes sustained in the population through reinfections, and it is true of dissipative systems in general.  Our measures may not only perform better in these cases, but also succeed in capturing dynamical properties not yet achievable by any other currently available means.  Because the interactions here are chosen randomly, there couldn't be any agent or social property that aligns with their role in the dynamics.  However, given the uniformly random interaction structure we still wish to understand the features that determine the sensitivity level of the disease's spread specifically because these are not features of the agents or of the underlying network structure -- these are features of the dynamics themselves.  Although we have made some encouraging progress on the problem of identifying and measuring these dynamical properties, this is still early work.  We expect to make further discoveries in the near future and look forward to collaborating with others on the exploration of temporal webs in its many varieties and applications.

\section*{Acknowledgements}
We would like to thank Koen Schoors for making this collaboration possible.
This work is supported by the Research Foundation Flanders (FWO-Flanders).

\bibliography{TemporalWebRefs}

\end{document}